%% file: manuscript.tex
\documentclass[fleqn,usenatbib]{mnras}
\usepackage{newtxtext,newtxmath}

\usepackage[T1]{fontenc}
\usepackage{graphicx}
\usepackage{dcolumn}
\usepackage{bm}

\usepackage[bottom]{footmisc}
\usepackage{calc}
\usepackage{color}
\usepackage{tikz}
\usetikzlibrary{positioning} 
\usepackage{xcolor}
\usepackage{overpic}
\usepackage{rotating}
\usetikzlibrary{arrows,shapes,chains}

\newcommand{\be}{\begin{equation}}
\newcommand{\ee}{\end{equation}}
\newcommand{\bs}{\begin{split}}
\newcommand{\es}{\end{split}}

\DeclareRobustCommand{\VAN}[3]{#2}
\let\VANthebibliography\thebibliography
\def\thebibliography{\DeclareRobustCommand{\VAN}[3]{##3}\VANthebibliography}

\usepackage{graphicx}	
\usepackage{amsmath}	

\title[Prospects and Strategies for Detecting GW BWM from SDSSJ1430$+$2303 by PTA]{Prospects and Strategies for Detecting Nonlinear Gravitational Wave Burst with Memory from Potential Merger Event SDSSJ1430$+$2303 by Pulsar Timing Arrays}

\author[J. W. Chen et al.]{
Jie-Wen Chen, Yiqiu Ma \thanks{Email: \href{myqphy@hust.edu.cn}{myqphy@hust.edu.cn}} and Yan Wang \thanks{Email: \href{ywang12@hust.edu.cn}{ywang12@hust.edu.cn}}
\\
$^{1}$Center for Gravitational Experiments, Hubei Key Laboratory of Gravitation and Quantum Physics, School of Physics, \\
Huazhong University of Science and Technology, Wuhan, 430074, P. R. China\\
$^{2}$Department of Astronomy, School of Physics, Huazhong University of Science and Technology, Wuhan, 430074, P. R. China
}

\date{Accepted XXX. Received YYY; in original form ZZZ}

\pubyear{2022}

\begin{document}
\label{firstpage}
\pagerange{\pageref{firstpage}--\pageref{lastpage}}
\maketitle

\begin{abstract}
The recently observed chirping  signature in the light curves of Seyfert 1 galaxy SDSSJ1430$+$2303 could be explained by a late-inspiralling supermassive binary black hole (SMBBH) system in the galactic center, which will merge in the near future (or could have merged already).
For the merging SMBBH scenario, SDSSJ1430$+$2303 can be a source of nonlinear gravitational wave (GW) burst with memory (BWM), which may provide a promising target for future pulsar timing array (PTA) observations.
In this work, we investigate the prospects for detecting the BWM signal from SDSSJ1430$+$2303 by the International PTA (IPTA) and FAST-PTA in the next $5$ years. We firstly propose strategies on searching for this target signal, including the selection of millisecond pulsars (MSPs) and the distribution of observation time.
Then we simulate PTA observations based on the proposed strategies and obtain the probability density functions of the network signal-to-noise ratio and parameter-estimation errors of the BWM signal, considering the uncertainties of parameters of the SMBBH and both white and red noises of the selected MSPs.
Our result shows that although IPTA can marginally detect the BWM in $5$ years, FAST-PTA can detect it with significantly higher confidence.
Moreover, the archived IPTA data is important in estimating the merger time of the SMBBH, when combined with the FAST-PTA data.
This work can serve as a guidance for future PTA observations and multi-messenger studies on SDSSJ1430$+$2303 and similar systems.
\end{abstract}




\section{Introduction}
\begin{figure}
  \centering
  \includegraphics[width=0.40\textwidth]{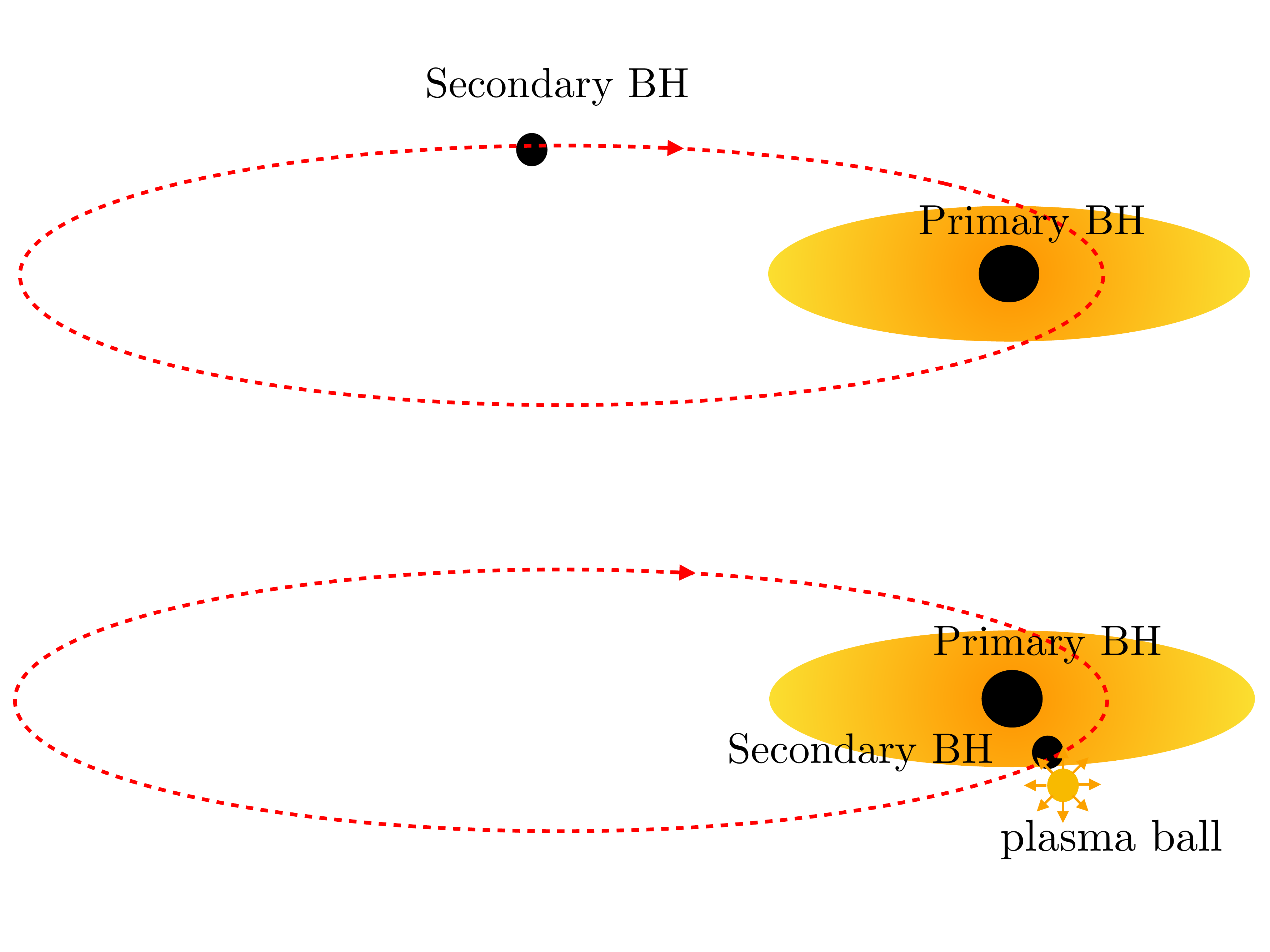}\\
  \caption{Schematic illustration of the proposed model for SDSSJ1430$+$2303, a highly eccentric supermassive binary black hole system with black hole masses $M_1\approx3\times 10^8M_{\odot},M_2\approx6\times10^7M_{\odot}$. The time varying light curves result from the electromagnetic radiation from the plasma ball brought by the collision of the secondary black hole with the accretion disk of the primary black hole.  }\label{fig:scheme}
\end{figure}

The recent observations of Seyfert 1 galaxy  SDSSJ143016.05$+$2303 (hereafter SDSSJ1430$+$2303) at various electromagnetic wavelengths, in particular at the optical and X-ray bands, reported periodically varying light curves with decreasing oscillation amplitude and period  \citep{Jiang:2022}. 
Although there could be other physical mechanisms (e.g., oscillations or instabilities of the AGN accretion disk \citep{Jiang:2022}), it is plausible that the observation signatures can be generated from a supermassive binary black hole (SMBBH) system in a highly eccentric orbit (see \autoref{fig:scheme}). 
Moreover, this SMBBH scenario provides the predictions that can be further tested by the near future multi-messenger observations, since the fitted orbit evolution models show that this SMBBH is currently at its late inspiral stage and will merge in a few years or could have merged recently. 
Despite this merger event will occur too soon to be captured by the future space-borne gravitational wave (GW) detectors (e.g., LISA \citep{LISA-1997, LISA-2017}, Taiji \citep{Taiji} and TianQin \citep{Luo_2016-TQ}) and also the frequency band of its GWs is too low to be detected by using the current ground-based GW detectors (e.g., LIGO \citep{Harry_2010, Aasi-2015} and Virgo \citep{Acernese_2014}), the nonlinear Christodulou GW memory effect \citep{Payne-1983, Blanchet-Damour-1992, Christodoulou-1991, Thorne-1992} can still have the chance to be detected by the pulsar timing arrays (PTAs) as pointed out by \citet{Jiang:2022}.

GW memory is a non-oscillatory contribution of the metric before and after the merger
event \citep{Favata-PRD-2009, Favata-ApJ-2009, vanHaasteren-2010, Pshirkov-2010}.
The non-linear GW memory contributed by the GW stress-energy tensor is a manifestation of the nonlinearity of the Einstein's gravitational theory. Detecting nonlinear GW memory effect can in principle provides a strong-field test of general relativity. The redshift of SDSSJ1430$+$2303 is estimated to be small ($\approx 0.08$), therefore the GW memory effect could be detected if SDSSJ1430$+$2303 is indeed an inspiralling SMBBH system.
In particular, since there will be a SMBBH merger event in a few years as predicted by the orbit evolution models of the SMBBH, it would be very important to test this model by detecting the corresponding nonlinear GW memory.

\citet{Jiang:2022} has pointed out that the GW burst with memory (BWM) effect is promising and can be detected by PTAs. The signal-to-noise ratio (S/N) is estimated to be $\approx 1$ based on a simple observation strategy -- biweekly observation of $20$ millisecond pulsars (MSPs) with a timing precision $100$ ns for a duration of $5$ years.
Hence it is important to give a prospective investigation for detecting the BWM signal by the recent-future PTA projects, considering more realistic performance of the radio telescopes and characteristics of the MSPs.
Currently, the global effort of PTA observation is carried out under the umbrella of the  International PTA (IPTA) consortium \citep{Hobbs_et_al_2010, Manchester_et_al_2013b}, which consists of four regional members --- the Parks PTA (PPTA) \citep{manchester_et_al._2013, Hobbs_2013}, the North American Nanohertz Observatory for GWs (NANOGrav) \citep{McLaughlin2013, Ransom_et_al_2019}, the European PTA (EPTA) \citep{Kramer&Champion2013} and the Indian PTA (InPTA) \citep{Joshi_et_al_2018}.
The IPTA has accumulated pulsar time of arrivals (TOAs) data for more than ten years,
offering 65 stable MSPs among which 15 have a timing precision below $1~\mu$s \citep{IPTA-DR2}.
Furthermore, the Five-hundred-meter Aperture Spherical radio Telescope (FAST) is now the largest single-dish radio telescope and is expected to provide unprecedented high precision measurement of pulsar time of arrivals  \citep{Nan_et_al_2011, Hobbs_et_al_2019}.
FAST will play an important role in the future Chinese PTA (CPTA) project \citep{Lee2016}.
The purpose of this work is to investigate the prospects and provide strategies for detecting the BWM signal from SDSSJ1430$+$2303 by IPTA observations and FAST-PTA  observations.

The rest of the article is organized as follows. In \autoref{sec:template},
we briefly introduce the BWM signal and the timing residual template induced by it.
In \autoref{sec:strategy}, we discuss the details of the observation strategies for detecting the BWM by PTA.
In \autoref{sec:simulation}, we analyze the PDFs of the network S/Ns and parameter-estimation errors, to test and further select the strategies, and also to investigate the prospects for detecting the GW signal.
This article is concluded in \autoref{sec:summary}.


\section{Timing Residual induced by the nonlinear GW memory}\label{sec:template}

GW memory effect, as briefly mentioned in the Introduction, is a non-oscillatory permanent distortion of the spacetime metric. 
This memory effect is caused by the DC changes in the time derivatives of the source multiple moments \citep{Favata-PRD-2009, Favata-ApJ-2009}.
The nonlinearity in the relativistic gravitational field equations indicates that the gravitational wave itself can carry time-varying energy momentum thereby can also contribute to the GW memory effect \citep{Thorne-1992, Favata-PRD-2009}. This phenomenon has been independently found by \citet{Payne-1983}, \citet{Blanchet-Damour-1992} and \citet{Christodoulou-1991}. It is intriguing and important since it can be used to test the general relativity in the strong-field regime and may be related to other topics relevant to the fundamental physics.
It has been proposed that the BWM effect can be detected by high precision measurement of the pulsar TOAs for a set of MSPs in a PTA \citep{vanHaasteren-2010, Pshirkov-2010}, and its upper limit has been improved over the years  \citep{2015ApJ...810..150A,2015MNRAS.446.1657W, 2020ApJ...889...38A}.

\begin{figure}
  \centering
  \begin{overpic}[width=0.40\textwidth]{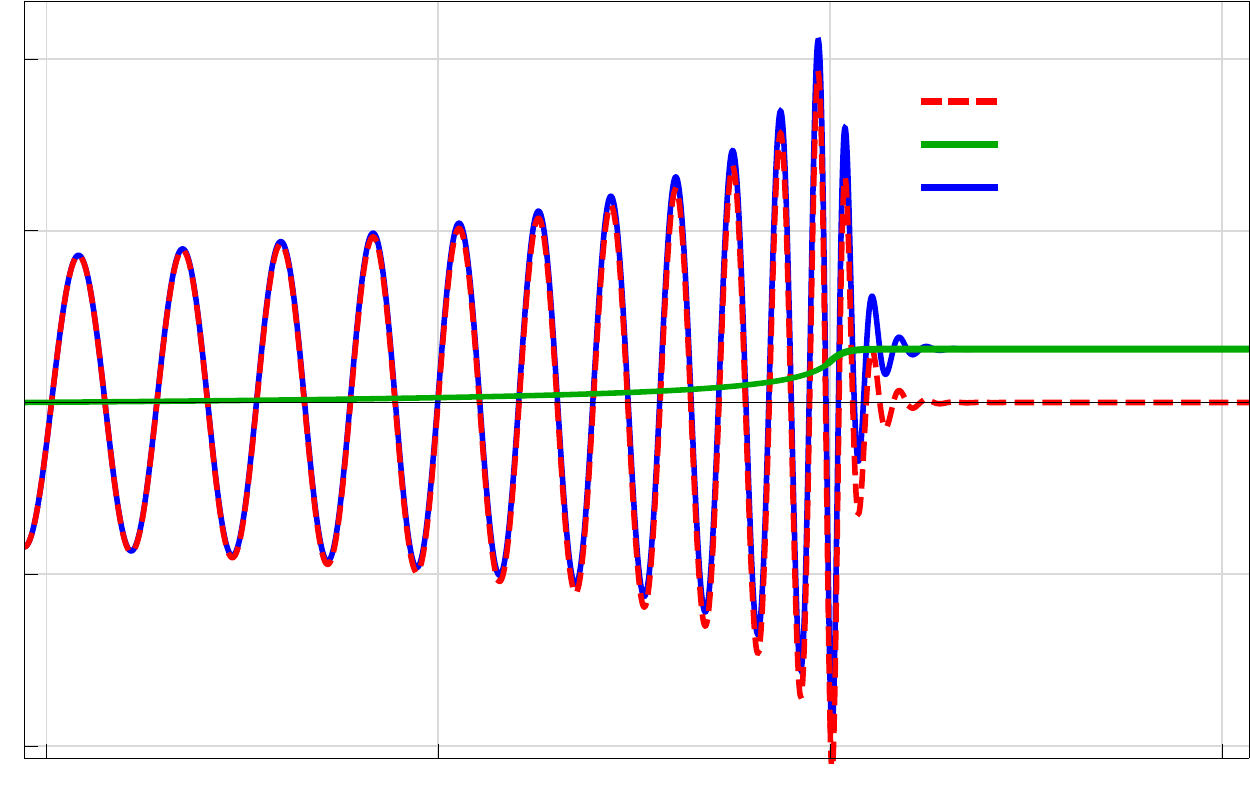}
  \put(42,-5){ $(t-t_m)$ / days}
  \put(-12,20){\begin{sideways} $(h_+-ih_\times) \times 10^{15}$ \end{sideways}}
  \put(81,55){\footnotesize oscillatory}
  \put(81,51.5){\footnotesize memory}
  \put(81,48){\footnotesize total}
  \put(0,-1){$-10$} \put(32,-1){$-5$} \put(66,-1){$0$} \put(96,-1){$5$}
  \put(-5,4){$-4$} \put(-5,17){$-2$} \put(-2,31){$0$} \put(-2,44){$2$} \put(-2,58){$4$}
  \end{overpic}\\
  ~\\
  \caption{Illustration for the waveform of the GW BWM from SDSSJ1430$+$2303.
  In this case, the masses of the two black holes are taken as $3 \times 10^8 M_\odot$ and $6 \times 10^7 M_\odot$ \citep{Jiang:2022}, $\iota$ is taken as $90^\circ$. The effective-one-body and quasi-normal mode waveforms are matched at the distance $3GM/c^2$ \citep{Favata-ApJ-2009}, with $M$ being the total mass of the SMBBH.
  }\label{fig:waveform}
\end{figure}
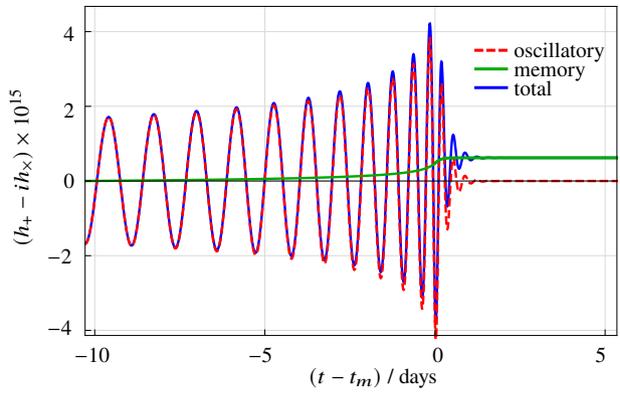

The DC GW memory is given by \citet{Favata-ApJ-2009, Favata-conference-2009}:
\be
    h^{\rm mem}_+ (t) =\frac{\sin^2\iota (17+\cos^2\iota)}{384 \pi R}\int^{t}_{-\infty}|I^{(3)}_{22}(t')|^2dt' \,,
\label{eq:gwmem}
\ee
where
we adopt a standard choice of polarization tensor in which $h^{\rm mem}_\times (t)=0$ \citep{Favata-ApJ-2009, Favata-conference-2009},
$I^{(3)}_{22}(t)$ is the third time derivative of the source mass moment for the $l=|m|=2$ mode,
$R$ is the luminosity distance between the source and the detector and $\iota$ is the inclination angle.
For the merger event of SDSSJ1430$+$2303, using the parameters estimated by \citet{Jiang:2022}, the GW memory effect around the merger time is plotted in \autoref{fig:waveform}. This waveform is obtained by matching the effective-one-body waveform (which extends the Post-Newtonian\,(PN) equations of motion to the plunge region by mapping the PN two-body description to a one-body problem in a parametrized deformed Schwarzschild  metric \citep{DAMOUR-EOB}) and the ringdown waveform  \citep[see][for details]{Favata-ApJ-2009, Favata-conference-2009, Berti-2006}.

The memory metric perturbation in \autoref{eq:gwmem} has a transient growing part which lasts for a timescale $ \big[1.45\,{\rm mHz}\times (2\times 10^6M_{\odot}/M) \big]^{-1}$ $\approx 1$  day \citep{Favata-ApJ-2009} (as shown in \autoref{fig:waveform}), where $M \approx 3 \times 10^8 M_\odot$ is the total mass of the SMBBH in SDSSJ1430$+$2303.
This is much shorter than the typical cadences ($2-4$ weeks) of PTA observations \citep{IPTA-DR2}.
Therefore we can used the Heaviside step function $\Theta$ to approximate the above memory waveform as
\be
h_+^{\rm mem}(t)=\Theta(t-t_m) h,
\ee
where $t_m$ is the merger time and $h$ is the strain amplitude of the GW memory.
Thus, for the $I_{\rm th}$ MSP in a PTA, the timing residual induced by the GW memory can be approximately written as \citep{Pshirkov-2010, vanHaasteren-2010}
\begin{align}
\label{eq:waveform-0}
s_I (t)=  (t-t_m) \Theta(t-t_m)F_I h,
\end{align}
where $F_I=(F_{I, +}~\cos 2\psi - F_{I, \times} ~\sin 2\psi)$ \citep{Taylor_2016}, $\psi$ is the GW polarization angle,
and
\be
\begin{split}
F_{I, +}=&\frac{1}{4(1-\cos \theta)} \left[ (1+\sin^2\delta)\cos^2\delta_I \cos(2\alpha-2\alpha_I) \right.  \\
&\left. -\sin 2\delta \sin 2\delta_I\cos(\alpha-\alpha_I)+\cos^2\delta(2-3\cos^2\delta_I)  \right],  \\
F_{I, \times}=&\frac{1}{2(1-\cos \theta)} \left[-\sin \delta \cos^2\delta_I \sin(2\alpha-2\alpha_I) \right.   \\
& \left. +\cos \delta \sin 2\delta_I\sin(\alpha-\alpha_I)  \right],
\end{split}
\ee
are the antenna pattern functions, with
$\cos \theta=\cos \delta \cos \delta_I \cos(\alpha-\alpha_I)+\sin \delta \sin \delta_I$. Here $\alpha=14^{\rm h}30^{\rm m}$ and $\delta=+23^\circ03'$ are the right ascension (RA) and declination (DEC) of the source,
and the $\alpha_I,\delta_I$ are the RA and DEC of the MSP, respectively.
The probability density functions (PDFs) of $t_m$ and $h$ are given by \citet{Jiang:2022}, where they consider two possible SMBBH 4.5PN orbit models that are obtained from the fits based on the optical observations (Model-1) and the optical + X-ray joint observations (Model-2). Model-1 predicts a expected merger time around 2023-2024, while Model-2 predicts that the merger event is likely to occur within several months from now.
For Model-2, if the initial observation epoch ($t_{I, i}$) is later than the merger time $t_m$, the template in \autoref{eq:waveform-0} should be modified to
\begin{align}
\label{eq:waveform}
s_I (t)= (t-t_{I, 0}) \Theta(t-t_m) F_I h,
\end{align}
where $t_{I, 0}={\rm \bf Max}[t_m, t_{I, i}]$.


\section{Observation strategy}\label{sec:strategy}

An optimal observation strategy for detecting the nonlinear BWM effect consists of a selection of MSPs and a distribution of observation time that maximize the network S/N
\begin{align}
\label{eq:SNR}
\rho^2 & ={\mathop \sum \limits_{I=1}^{N_P}} \rho_I^2
=  {\mathop \sum \limits_{I=1}^{N_P}} {\mathop \sum \limits_{n=1}^{N_I}}  \frac{ s_I^2(t_n)}{\sigma_{I, W}^2+\sigma_{I, R}^2(t_n)}  \nonumber \\
& \approx  {\mathop \sum \limits_{I=1}^{N_P}} \int^{t_{I, i}+T_I}_{t_{I, i}} \frac{dt}{\Delta t_I} \frac{ s_I^2(t)}{\sigma_{I, W}^2+\sigma_{I, R}^2(t)} \,.
\end{align}
Here, the $N_P$ is the number of MSPs in the PTA,
$\rho_I$ is the individual S/N for the $I_{\rm th}$ MSP.
$\sigma_{I, W}$ and $\sigma_{I, R}$ are the root-mean-square (r.m.s.) of white noise and effective r.m.s. of red noise \citep{Shannon_2010}, respectively.
$t_{I, i}$, $T_I$ and $\Delta t_I$ are the initial date, span and cadence of the observation, respectively.
For simplicity, we assume that the cadence is a constant $\Delta t_I = T_I/(N_I-1)$ in this work, where
$N_I$ is the total number of data points for this MSP.

The noises $\sigma_{I, W}$ and $\sigma_{I, R}$
will be discussed in \autoref{sec:noise}.
The details of the observation strategy will be demonstrated in \autoref{sec:strategy-design} and the results will be discussed in \autoref{sec:Preliminary-Strategies}.

\subsection{Noise Evaluation}\label{sec:noise}

\input{msptable.tex}

$\bullet$ \textbf{White noise}\,---
The IPTA DR2 \citep{IPTA-DR2} has released $\sigma_{I, W}$ for 65 MSPs, as is listed in column 2 of \autoref{Tab:white-noise}, some of which are within the sky area accessible by FAST
($-14^\circ< {\rm DEC} <66^\circ$).
The white noise r.m.s. $\sigma_{I, W}$ for these MSPs, when observed by FAST, can be estimated as follows.

For a specific MSP, the white noise consists of the following jitter noise ($\sigma_{I, j}$) and radiometer noise ($\sigma_{I, r}$) \citep{Shannon_2012, Hobbs_et_al_2019},
\be
\label{eq:white-noise}
\begin{split}
\sigma_{I, j} =0.2 W_I \sqrt{\frac{P_I}{\Delta \tau_I}} \,; \quad
\sigma_{I, r} =\frac{W_I T_{\rm sys}}{G S_I \sqrt{2\Delta f \Delta \tau_I}}\sqrt{\frac{W_I}{P_I-W_I}} \,.
\end{split}
\ee
The total white noise $\sigma_{I, W}=\sqrt{\sigma_{I, j}^2+\sigma_{I, r}^2}$.
Here, the $P_I$, $W_I$ and $\Delta \tau_I$ are the spin period, the pulse width and the integration time of each data point, respectively. 
$S_I$ is the flux density of the MSP. 
In this work, $\Delta \tau_I$ is simply assumed to be a constant for a specific MSP, so that the total integration time allocated to the MSP is $\tau_I = N_I \Delta \tau_I$.
The $G,T_{\rm sys},\Delta f$ are the gain,
the system temperature,
and the receiver's bandwidth of the telescope, respectively. The corresponding values for FAST is listed in \autoref{tab:FAST}. 
As in \citet{Hobbs_et_al_2019}, we ignore the dependence of $G$ and $T_{\rm sys}$ on the zenith angle and simply set $G=16.0$ K Jy$^{-1}$ and $T_{\rm sys}=20$ K in this work.
\autoref{eq:white-noise} implies that $\sigma_{I, W} \propto \Delta \tau_I^{-1/2}$,
and in practical measurements, $\Delta \tau_I$ is required to be not smaller than $5$ min \citep{1995ApJ...452..814R} and is typically $\approx 10-30$ min .
To calculate the r.m.s. of white noises for those MSPs by FAST,
we also list the MSP parameters  $S_{I}$, $P_I$ and $W_{I}$
in columns 3-5 of  \autoref{Tab:white-noise}, and show the resulting white noise for $\Delta \tau_I=5$ min in column 6.

\begin{table}
\centering
\caption{The telescope instrumental parameters for FAST  (\citep{Jiang_2020}; \href{https://fast.bao.ac.cn/cms/article/97/}{https://fast.bao.ac.cn/cms/article/97/}). }
\label{tab:FAST}
\begin{tabular}{|c|c|}
\hline
 parameters & values \\
\hline
$G$  & $11.0-16.0$ K Jy$^{-1}$  \\
$T_{\rm sys}$ & $19-27$ K \\
band & $1.4$ GHz \\
$\Delta f$ & $400$ MHz\\
\hline
\end{tabular}
\end{table}

$\bullet$ \textbf{Red noise}\,---
The time correlated pattern observed in timing residuals is usually referred to as spin noise or timing noise. It has a red power spectral density and can arise from the spin irregularity of the pulsar that may be caused by the varying coupling between the crust and the internal core of the neutron star \citep{2015JPhCS.610a2019W}. The typical strength of the timing noise in the MSP is orders of magnitude weaker than the ones in the normal pulsar or magnetar. However, it can appear in the long term high precision measurements of TOAs of MSPs and has adverse impact on the detection of GW by delaying the expected detection time \citep{2013CQGra..30v4015S}. As the reason that will become clear later, adding red noise intrinsic in the MSPs in our analysis for FAST is crucial for obtaining a realistic observation strategy.
Here we adopt the model to estimate the level of the red noise given by \citet{Shannon_2010}:
\be
\ln \frac{\sigma_{I, R}}{1 {\rm \mu s}} \in N\left[ \ln \frac{\tilde  \sigma_{I, R}}{1 {\rm \mu s}}, \delta^2 \right] \,,
\ee
where
\be
\frac{\tilde \sigma_{I, R}}{1 {\rm \mu s}}=C \left(\frac{P_I}{1 {\rm s}}\right)^{-\alpha-2\beta} \left(\frac{\dot P_I}{10^{-15}}\right)^\beta \left(\frac{t-t_{I, i}}{1 {\rm yr}}\right)^\gamma \,,
\ee
with $\ln C \in {N[-1.2, 1.25^2]}$, $\alpha \in N[-0.9, 0.2^2]$, $\beta \in N[0.8, 0.05^2]$, $\gamma \in N[2.3, 0.15^2]$ and $\delta \in N[1.1, 0.05^2]$ \citep{Lam_2016}.
We find that for the observations with $t-t_{I, i} \leq 10$ years, the above model can be simplified as
\be
\label{Eq:red-noise}
\sigma_{I, R}(t) \in \hat \sigma_{I, R} ~ 10^{N[0, \delta_{I}^2]}=\hat \sigma_{I, R, 1{\rm~yr}} \left( \frac{t-t_{I, i}}{\rm 1~ yr} \right)^{2.3} 10^{N[0, \delta_{I}^2]} \,,
\ee
where
$\hat \sigma_{I, R}$ is the expectation value of the red noise,
and $\delta_I$ is the parameter of uncertainty.
The expected $1$-year red noise $\hat \sigma_{I, R, 1{\rm~yr}}$ and the parameter $\delta_I$ for the MSPs are listed in columns 8-9 of \autoref{Tab:white-noise}.

\subsection{MSP selection and observation time allocation  }\label{sec:strategy-design}

For each selected MSP,
the observation time to be determined includes 4 free  parameters: $T_I$, $t_{I, i}$, $\tau_I$ and $\Delta t_I$ (or alternatively $\Delta \tau_I$, since $\Delta \tau_I=\tau_I /(T_I/\Delta t_I+1)$~).
Note that designing the observation strategy concerns a complicated process.
The details of the process can be summarised as the following steps.

\textbf{Step 1-\,Selection of $\Delta \tau_I$}\,---
The integration time of each data point $\Delta \tau_I$ affects $\rho_I$, as well as the network S/N $\rho$, by determining the white noise level of the MSP.
An optimal strategy requires a large $\rho_I^2/\tau_I$, which is
\be
\label{eq:rho/tau}
\frac{\rho_I^2}{\tau_I}=\frac{1}{N_I} {\mathop \sum \limits_{n=1}^{N_I}}  \frac{ s_I^2 (t_n)}{5
~{\rm min}\cdot\sigma_{I, W, 5{\rm min}}^2+\sigma_{I, R}^2 (t_n) \Delta \tau_I} \,,
\ee
according to \autoref{eq:white-noise}, where $\sigma_{I, W, 5{\rm min}}$ is the white noise with $\Delta \tau_I=5$ min.
To yield a maximum $\rho_I^2/\tau_I$ in the presence of red noise,
$\Delta \tau_I$ should contain the minimum value $5$ min.
Therefore,
in the following analyses, $\Delta \tau_I=5$ min will be adopted as a preliminary selection,
then its adjustment will be discussed in \textbf{Step 6}.

\textbf{Step 2-\,Distribution of $T_I$}\,---
Generally speaking, $\rho_I$ grows as the observation span $T_I$ increases. However, at the late observation stage when the red noise dominates, the S/N will be saturated. This fact can be seen by substituting  \autoref{eq:waveform} into \autoref{eq:SNR}, which leads to:
\begin{align}
\rho_I^2 \propto \int^{t_{i,I}+T_I}_{t_{i,I}} dt \frac{(t-t_{I, 0})^2 \Theta (t-t_m)}{\sigma_{I, W}^2+\sigma_{I, R}^2(t)}.
\end{align}
As mentioned above, $\sigma_{I, W}$ is independent from the span, while  $\sigma_{I, R}(t) \propto (t-t_{I, i})^{2.3}$, if we estimate $\sigma_{I, R}$ as its expected value $\hat \sigma_{I, R}$ in \autoref{Eq:red-noise}. Therefore, one can expect that:
at the early observation stage,
the white noise dominates over the red one and the S/N increases as the span $T_I$ becomes larger;
at the late stage, the red noise dominates and
the S/N converges.
To characterise this S/N convergence at the late stage, we define the S/N growing time scale
\be
\label{eq:span-limit}
T_I^{*} =1~{\rm yr} \times \left(\frac{3 \sigma_{I, W, {\rm 5 min}}^2}{1.6 \hat \sigma_{I, R, {\rm 1 yr}}^2} \right)^{\frac{1}{4.6}},
\ee
by  $\sigma^2_{I, W, 5~{\rm min}} \sim \sigma^2_{I, R}(t)$ at $t=T_I^*+t_{I, i}$. 
This result is based on the simple assumption that $h^{\rm mem}_{+}$ is non-zero at the beginning of the observation (i.e. $t_{I, i} \geq t_{m}$).

In a $5$-year PTA observation,
the span for a specific MSP $T_{I}$ should be designed to be no larger than $T_I^{*}$ (i.e. $T_{I}={\rm \bf Min}[T_I^{*}, 5~{\rm yr}]$),
otherwise the observation during the excess span has little contribution to $\rho_I$.
The values of $T_I^{*}$ for FAST-PTA and IPTA are listed in columns 10 and 11 of \autoref{Tab:white-noise},
where we assume that the IPTA white noises in column 2 are measured with $\Delta \tau_I=20$ min for all MSPs for simplicity.

\textbf{Step 3-\,Selection of $t_{I, i}$}\,---
To probe the BWM signal from SDSSJ1430$+$2303, the PTA observation is expected to start as early as possible,
hence we take $t_{I, i}=2022.3$ (mid-April of the year 2022) for most MSPs.
However, for some specific MSPs, $t_{I, i}$ should be chosen differently for the following reason.
In the Model-1 given by \cite{Jiang:2022},
the merger time $t_m$ may be later than the date $2023.4$ with a considerable probability.
For the MSPs with $T_{I}^{*} \lesssim 1$ year ( e.g. J1939$+$2134 with $T_{I}^{*}=1.1$ yr in FAST-PTA, as shown in \autoref{Tab:white-noise}),
it is possible that there are no BWM signals during most of the spans, if $t_{I, i}=2022.3$ is taken.
Therefore, we should delay the initial dates of observation for these MSPs.

In general, two cases are desired in observations:
(i) $t_m$ resides in the observation time range $[t_{I, i}\, ,\, t_{I, i}+T_I]$, so that it can be measured (see \autoref{sec:PEE} for details);
(ii) $t_{I, i}>t_m$, i.e. the BWM signal is non-zero initially, which leads to the largest $\rho_I$.
Therefore, $t_{I, i}$ should be selected that both the cases can happen with significant probabilities.
As a result, for the MSPs with $T_I^*\lesssim 1$ yr, $t_{I, i}$ is suggested to be around the date $2023.4$, which is the moment that $t_m$ takes the peaked value of PDF in Model-1 \citep{Jiang_2020}.
Finally, we notice that the above selection of $t_{I, i}$ will not affect $\rho_I$ in Model-2, in which the case (ii) above is satisfied with probability $\approx 1$.

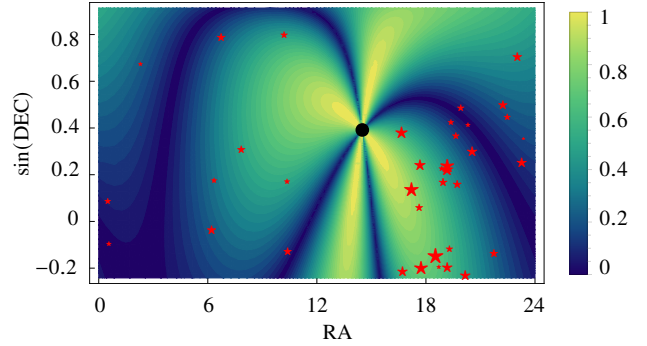
\begin{figure}
  \centering
  \begin{overpic}[width=0.4\textwidth]{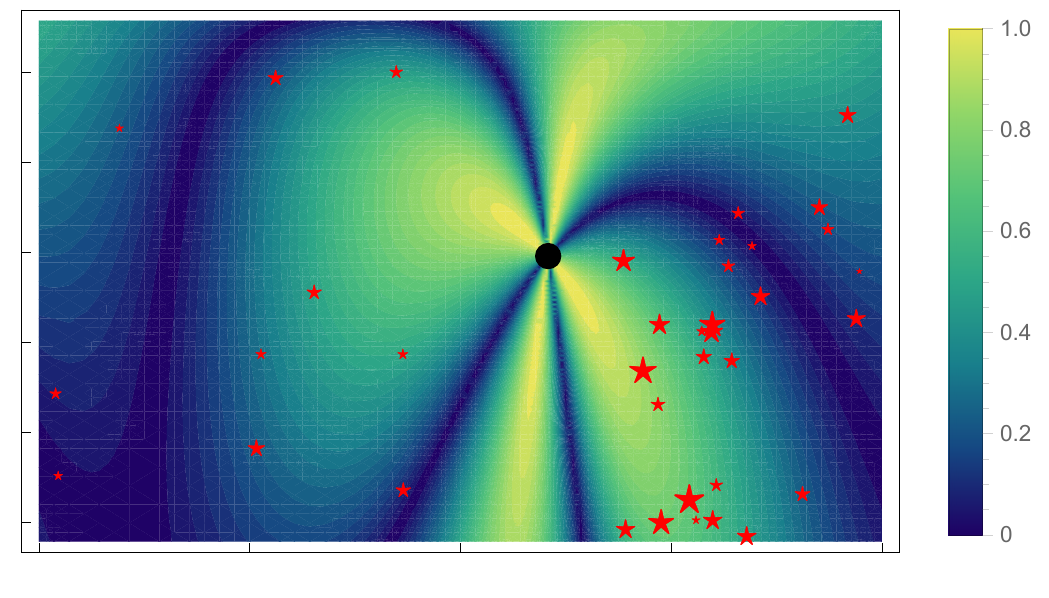}
  \put(45,-6){RA}
  \put(-12,24){\begin{sideways} $\sin ({\rm DEC})$ \end{sideways}}
  \put(3,-1){$0$} \put(23,-1){$6$} \put(42,-1){$12$} \put(62,-1){$18$} \put(82,-1){$24$}
  \put(-8,6){$-0.2$} \put(-2,15){$0$} \put(-5,24){$0.2$} \put(-5,32){$0.4$} \put(-5,40){$0.6$} \put(-5,48){$0.8$}
  \put(94,33){\textcolor{white}{\Huge $\blacksquare$}} 
  \put(94,5){\textcolor{white}{\Huge $\blacksquare$}}
  \put(94,14){\textcolor{white}{\Huge $\blacksquare$}}
  \put(94,24){\textcolor{white}{\Huge $\blacksquare$}}
  \put(94,42){\textcolor{white}{\Huge $\blacksquare$}}
  \put(94,51){\textcolor{white}{\Huge $\blacksquare$}}
  \put(96,52){$1$} \put(96,6){$0$} \put(96,15){$0.2$} \put(96,24){$0.4$} \put(96,33){$0.6$} \put(96,42){$0.8$}
  \end{overpic}\\
  ~\\
  \caption{Illustration for the antenna pattern function ($|F_{I}|$) of MSPs in the FAST sky.  
  Here $|F_{I}|$ is illustrated by the brightness, and the brighter region represents larger $|F_{I}|$, hence stronger signal. 
  The sky locations of MSPs are marked by red stars  
  with sizes referring to the noise levels. 
  Larger stars means MSPs with smaller noise  $\sqrt{\sigma_{I, W}^2+\sigma_{I, R}^2}$.  
  The black point marks the location of SDSSJ1430$+$2303.
  In the presented case, the noise level is evaluated based on FAST-PTA,
  assuming $\psi=22.5^\circ$, $\sigma_{I, R}=\hat \sigma_{I, R}$, and $t-t_{I, i}=5$ yr}\label{Fig:sky}
\end{figure}

\input{best_msp.tex}

\textbf{Step 4-\,MSP selection}\,---
The  $\rho_I$ depends on the antenna patter function,  sky location and noise level of the MSP. As an example, in \autoref{Fig:sky}, we plot the antenna patter function and the sky locations of the MSPs in the FAST sky,
and make the noise inverse to the size of the stars.
In the following, we will select MSPs considering these factors.
Supposing all MSPs are observed during their efficient spans $T_I \leq T_I^{*}$, a MSP with a larger $\rho_I^2/\tau_I$ should have a higher priority to be chosen.
Considering the uncertainties of $\psi$ and $\sigma_{I, R}$,
we assume $\psi \in U[0, \pi]$ and take $\sigma_{I, R}=\hat \sigma_{I, R}$, then we define the priority parameter $p_I$ for MSP selection as:
\begin{align}
\label{Eq:priority}
p_I=
\frac{F_{I, +}^2+F_{I, \times}^2}{N_I} {\mathop \sum \limits_{n=1}^{N_I}} \frac{(t_n-t_{I, 0})^2 \Theta(t_n-t_m)}{\sigma_{I, W}^2+\hat \sigma_{I, R}^2}  \propto \frac{\hat \rho_I^2}{\tau_I},
\end{align}
where $\hat \rho^2$ is the $\psi$-averaged  $\rho^2$. For a specific telescope, the MSP with a larger $p_I$ has a greater priority in observation. $p_I$ computed and assigned to each MSPs for both FAST-PTA and IPTA are listed in \autoref{tab:strategy}.

\textbf{Step 5-\,Distribution of $\tau_I$}\,--- For a fixed  total integration time $\tau={\mathop \sum \limits_{I=1}^{N_P}} \tau_I$, different distribution strategies of $\tau_I$ among all MSPs will yield different values of $\rho$ and different distributions of $\rho_I$.
The distribution strategy should be taken
in the way to avoid the situation that the network S/N is dominated by one or a few MSP(s). In this work, we distribute $\tau_I$ to ensure that the aforementioned $\psi-$averaged $\rho_I^2$ for all MSPs are the same. In another words, \autoref{Eq:priority} means that the distribution strategy is $\tau_I \propto p_I^{-1}$.
\autoref{tab:strategy} lists the values of
$\tau_I/{\rm \bf Min}[\tau_I]$, which represents the distribution ratio of $\tau_I$ for a MSP with respect to the MSP with largest $p_I$.

\begin{figure*}
    \centering
   \input{flow.tex}
    \caption{Process of designing the observation strategy. In particular, the details of \textbf{Step 6} are presented.}
    \label{fig:flow}
\end{figure*}
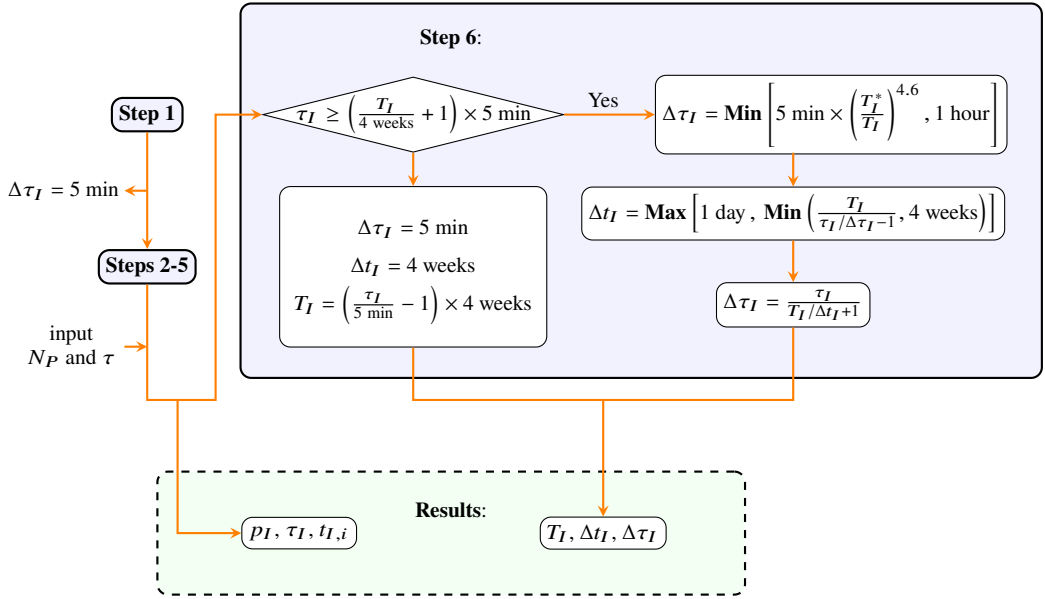

\textbf{Step 6-\,Further adjustments }\,---
The above analyses are based on the minimum integration time  $\Delta \tau_I=5$ min for each data point.
A small $\Delta \tau_I$ corresponds to a high-cadence observation (small $\Delta t_I$) at a fixed $\tau_I$,
which means complicated telescope time arrangements and should be avoided in practice.
Hence, we will take a further selection of $\Delta \tau_I$ and $\Delta t_I$,
which may lead to adjustments of other parameters in the strategy.

The details of the further selection are shown in \autoref{fig:flow}, with the following factors considered.
Firstly, both $\Delta \tau_I \geq 5$ min and $\Delta t_I \leq 4$ weeks \citep{IPTA-DR2} are required in PTA observations, which leads to the lower limit $\tau_I \geq 5~{\rm min} \times \left[T_I/(4~{\rm weeks})+1 \right]$.
If $\tau_I$ given by \textbf{Step 5} is smaller than this lower limit, the parameters $T_I$ or $\tau_I$ should change in order to satisfy the above requirements of $\Delta \tau_I$ and $\Delta t_I$.
Here we re-select $T_I$ at $\Delta \tau_I=5$ min and $\Delta t_I=4$ weeks.
Secondly, the value of $\Delta \tau_I$ is expected to have an upper limit,
otherwise the red noise will dominate during the observation span $T_I$.
The upper limit equals $5$ min $\times (T_I^*/T_I)^{4.6}$ according to \autoref{eq:span-limit}.
Furthermore, the configuration of FAST indicates that $\Delta t_I \geq 1$ day and $\Delta \tau_I \leq 1$ hour \citep{Jiang_2020}.
Since IPTA contains multiple telescopes, the above constraints of $\Delta t_I$  and $\Delta \tau_I$ can be removed \citep{Yi2014, Dolch_2016, Perera2019}, but in reality they hold for most cases.
Therefore, we adopt the constraints for both FAST-PTA and IPTA in this step.
Given above, the parameters $\Delta t_I$, $\Delta \tau_I$ and $T_I$ are reset in light of the process given in \autoref{fig:flow}.
In our strategy, the constraints $\Delta \tau_I \geq 5$ min and $1$ day $\leq \Delta t_I \leq 4$ weeks are stringently satisfied, while the other ones above are not.

Finally, we emphasize that if all the parameters $T_I$, $t_{I, i}$, $\tau_I$, $\Delta \tau_I$ and $\Delta t_I$ are chosen, the observation epochs of all data points for the MSP are determined.

\subsection{Results} \label{sec:Preliminary-Strategies}

Following the above \textbf{Steps 1-5}, we can choose MSPs with high priority parameters $p_I$, and obtain their parameters $T_I$, $t_{I, i}$ and $\tau_I/{\rm \bf Min}[\tau_I]$, for general cases with various $\tau$.
In \autoref{tab:strategy}, 
we list the selected 20 MSPs with highest $p_I$ for $5$-year FAST-PTA and IPTA started at $2022.3$,
as well as their parameters. 

\begin{figure*}
  \centering
\input{duty-cycle.tex}\\
  ~\\~\\
  \caption{Illustration of the designed observation epochs for the strategy using the 7 MSPs with highest priorities in \autoref{tab:strategy} for the  FAST-PTA, with the total integration time $\tau=500$ hours.} \label{Fig:duty-cycle}
\end{figure*}
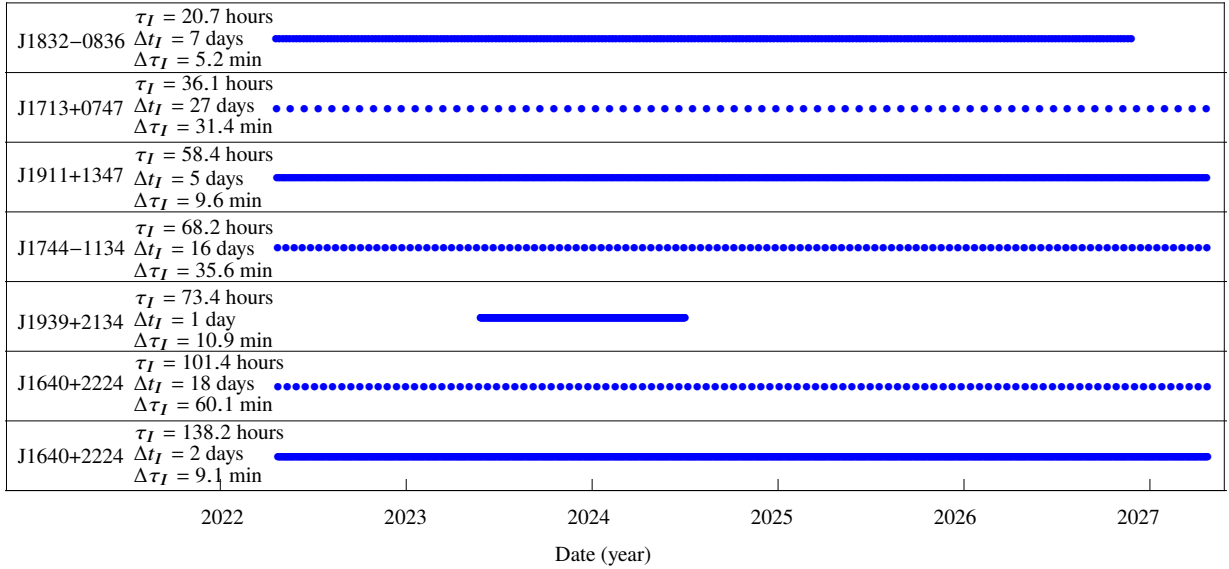

In practical PTA observations, only part of the MSPs in \autoref{tab:strategy} are needed to detect the BWM signal.
If the MSPs are selected and $\tau$ is fixed,
the observation epochs of all data points are obtained by the process in \autoref{fig:flow}.
As an example, for the strategy using 7 MSPs with the highest priorities for a $5$-year FAST-PTA in \autoref{tab:strategy} and with $\tau=500$ hours,
the designed observation epochs are illustrated in \autoref{Fig:duty-cycle}.

Furthermore, since \textbf{Step 4} only tells us to choose MSPs with the highest $p_I$ in \autoref{tab:strategy}, the number of the MSP is still a free parameter in the strategy. In the next section, we will figure out the MSP number and thereby further select the strategies.


\section{Further Selection and Tests of Strategies by Simulations}\label{sec:simulation}

In this section,
we will further select and test the strategies and investigate the prospects for detecting the BWM signal by IPTA and FAST-PTA, respectively.
Since each strategy includes $N_P$ MSPs with the highest priorities in \autoref{tab:strategy}, it will be denoted as the "best-$N_P$" strategy hereafter.
One goal of this part is to fix the value of $N_P$ for the most recommended strategy.

In \autoref{sec:strategy},
the uncertainties of $t_m$, $h$, $\psi$ and $\sigma_{I, R}$ are not considered in details.
Therefore, we will simulate the GW BWM signal and red noise by  taking the PDFs of $t_m$ and $h$ given in \cite{Jiang:2022} (for both Model-1 and Model-2) as well as
the PDF of $\sigma_{I, R}$ given in \autoref{Eq:red-noise}
and $\psi \in U[0, \pi]$.
Our simulations will be shown on two aspects --- the network S/Ns and the parameter-estimation errors (PEEs).

\subsection{Network S/Ns: Further Selection of Strategies \&. Prospects for Detecting BWM Signal}

\begin{figure*}
  \centering
  \begin{overpic}[width=0.9\textwidth]{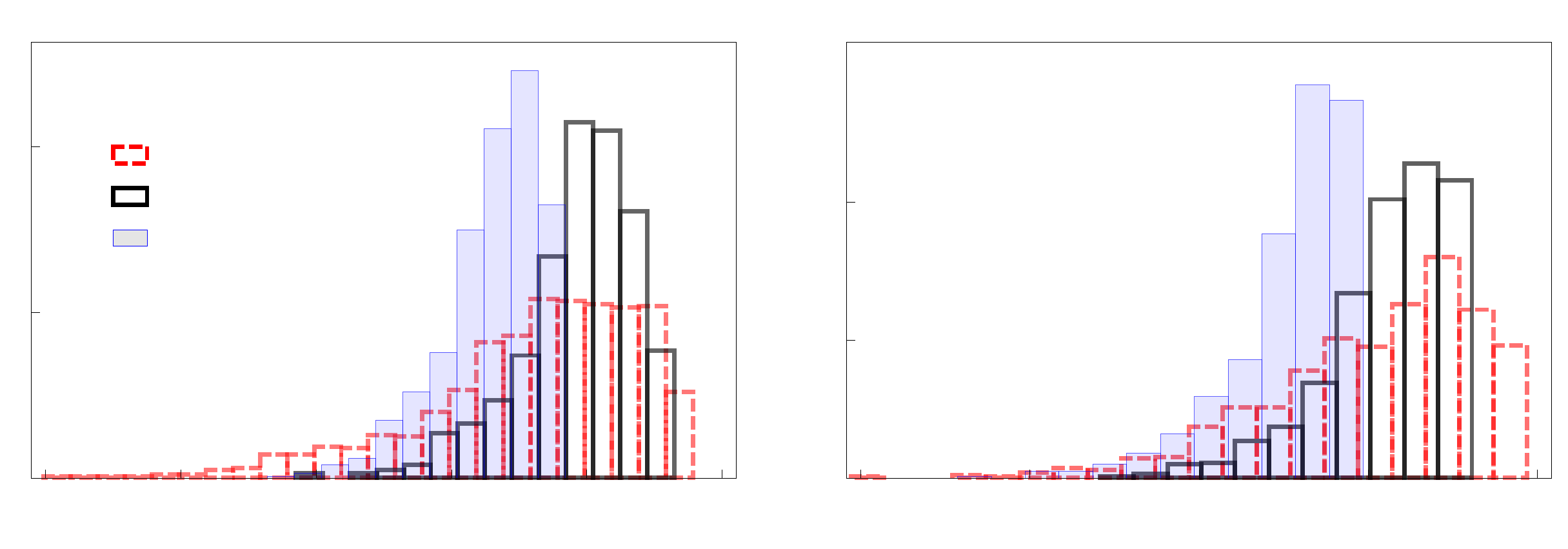}
  \put(3,2.5){$-3$} \put(10,2.5){$-2$} \put(18,2.5){$-1$} \put(28,2.5){$0$} \put(37,2.5){$1$} \put(45,2.5){$2$}
  \put(53,2.5){$-2$} \put(64,2.5){$-1$} \put(76,2.5){$0$} \put(87,2.5){$1$} \put(96,2.5){$2$}
  \put(-1.5,15){$100$}  \put(-1.5,25){$200$}
  \put(-4,15){\begin{sideways} Counts \end{sideways}}
  \put(49,15){\begin{sideways} Counts \end{sideways}}
  \put(50.5,13){$100$}  \put(50.5,22){$200$}
  \put(10.5,19){best-$20$}  \put(10.5,22){best-$5$} \put(10.5,24.8){best-$1$}
  \put(22,0){$\log_{10} \rho$}
  \put(74,0){$\log_{10} \rho$}
  \end{overpic}\\
  \caption{PDFs of network S/Ns $\rho$ given by $1000$ realizations, for a $5$-year FAST-PTA observation with $\tau=1000$ hours. The results for Model-1  are shown in the left panel and the ones for Model-2 are shown in the right panel.}\label{Fig:SNR}
\end{figure*}
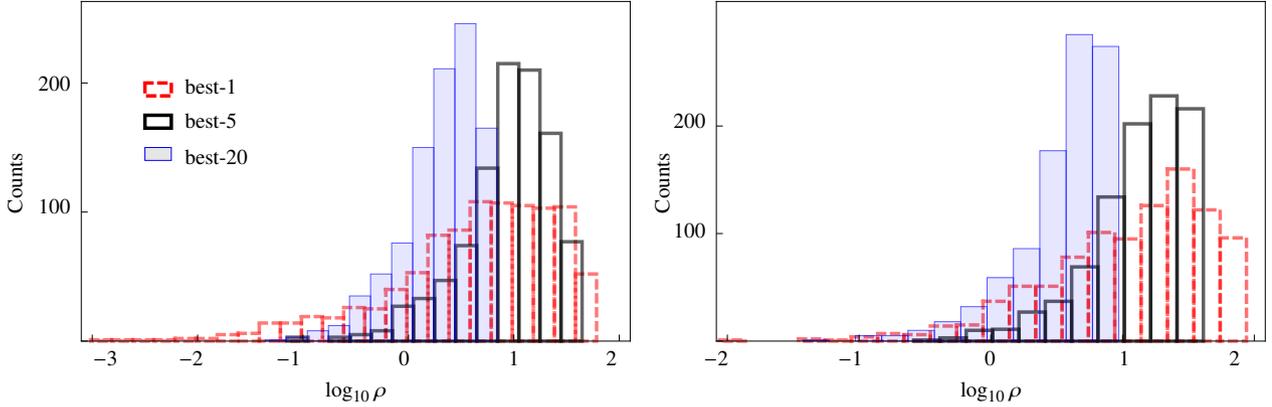

In the following simulations, we select among the best-$N_P$ strategies with $1\leq N_P \leq 20$.

We plot the PDFs of network S/Ns in \autoref{Fig:SNR} for the best-1, best-5 and best-20 strategies in FAST-PTA. Moreover, the averaged values of $\rho$ (actually $\sqrt{\langle\rho^2\rangle}$) for more various strategies are listed in \autoref{tab:PDF}.
The results exhibit two features.
Firstly,
the PDF profiles of $\rho$ are narrower as $N_P$ becomes larger.
This is because $\sigma_{I, R}$ in our treatment is uncorrelated among different MSPs,
and including more MSPs can reduce the uncertainties of $\rho$ due to the individual red noises.
Secondly, the averaged $\rho$ decreases as $N_P$ increases (for a fixed $\tau$).
This is due to the fact that including more MSPs means allocating $\tau$ to more MSPs with lower priorities. 

However,
it does not mean that the strategies with larger $N_P$ are less favored, because the above features are not sufficient to judge the strategies.
In practical observations,
a significant signal requires $\rho$ to be larger than a threshold, which is taken to be $8$ in this work \citep{Taylor_2016}.
Therefore, the probability of $\rho>8$ (denoted as $p(\rho>8)$ ), instead of the averaged $\rho$, is the key figure to judge strategies, and we seek for strategies with large $p(\rho>8)$.

Additionally, recall that in  \autoref{sec:strategy-design}, we expect $\rho_I$ for various MSPs to have equal values.
However, note that this is nearly impossible in practical observations, due to the uncertainties of parameters $\psi$, $t_m$ and $\sigma_{I, R}$.
For strategies with inhomogenous $\rho_I$, the network S/N will be dominated by a few MSPs and may easily biased by the individual noises of these MSPs \citep{Chen-Wang-2022}.
To quantify the inhomogeneity of $\rho$, we introduce the parameter
\be
\lambda =\frac{{\rm \bf Max}[\rho_I^2]}{\rho^2},
\ee
which takes the value between $N_P^{-1}$ and $1$. A smaller $\lambda$ represents a low $\rho$-inhomogeneity.
Here we adopt the criterion $\lambda<0.5$,
and the probability of $\lambda<0.5$ (denoted as "$p(\lambda<0.5)$") can be taken as another important parameter to judge strategies.
In details, among the strategies with almost the same $p(\rho>8)$,
the one with the largest $p(\lambda<0.5)$ should be the most recommended one.

Our results are summarised as follows.
In \autoref{tab:PDF}, we list the resulting $p(\rho>8)$ and $p(\lambda<0.5)$ for various strategies in IPTA and FAST-PTA.
Firstly, the largest values of $p(\rho>8)$ for FAST-PTA are $\{0.48, 0.61, 0.73\}$ for orbit Model-1 and $\{0.71, 0.82, 0.88\}$ for Model-2 with $\tau=\{500, 1000, 2000\}$ hours in $5$ years, respectively.
Meanwhile, the probability $p(\rho>8)$ for IPTA is significantly smaller than the FAST-PTA result at the same condition.
This means that a $5$-year FAST-PTA observation has considerable potential to detect the GW BWM signal from SDSSJ1430$+$0323,
while the IPTA is marginally capable to detect this signal in $5$ years, with total integration time $\tau \leq 2000$ hours.
Secondly, the largest $p(\rho>8)$ for FAST-PTA appear at $N_P=3-4$ in Model-1 and at $N_P=3-6$ in Model-2.
Furthermore, there are other strategies with $p(\rho>8)$ slightly smaller than the largest values (with differences below $10\%$), which are also acceptable.
For example, in the case $\tau=1000$ hours, the recommended strategies include those with $N_P= 2-6$ for Model-1 and $N_P= 2-8$ for Model-2.
For a general consideration with both Model-1 and Model-2, the strategies with $N_P= 2-6$ are recommended.
Moreover, it is clear that $p(\lambda<0.5)$ increases significantly as $N_P$ grows.
Therefore, in light of both $p(\rho>8)$ and $p(\lambda<0.5)$,
the most recommended strategy at $\tau=1000$ hours is the best-$6$ one.

\input{statistical_signature.tex}

\subsection{Parameter-estimation Errors } \label{sec:PEE}

In this part,
we will test the capability of the recommended best-6 strategy in parameter estimation.

The parameters of the GW BWM signal can be included in a
vector $\{ \lambda^i \}= \{ h, \psi, t_m \}$ with $i=1, 2, 3$,
and the variance of the PEE $\lambda^i$ can be evaluated as
$(\Delta \lambda^i)^2=(F^{-1})^{ii}$,
where $(F^{-1})^{ij}$ is the inverse of the Fisher information matrix:
\begin{align}
F_{ij}
= {\mathop \sum \limits_{I, n}} \frac{\partial s_I(t_n)}{\partial \lambda^i} \frac{\partial s_I (t_n)}{\partial \lambda^j} \frac{1}{\sigma_{I, W}^2+\sigma_{I, R}^2(t_n)}  \,.
\end{align}
In general, one expects that the Fisher information matrix is three dimensional and contains a positive determinant (det$[F_{i j}]>0$).
However, if the merger time $t_m$ is earlier than the initial observation epoch $t_{I, i}$, \autoref{eq:waveform} will yield $\partial s_I/\partial t_m=0$, which means $t_m$ cannot be measured at all or equivalently $\Delta t_m=+\infty$.
In this case, we use the two dimensional Fisher information matrix $\mathcal F_{i j}=F_{i j}$ (with $i, j = 1, 2$), and its inverse matrix is denoted as $(\mathcal F^{-1})^{ij}$.
Given above, we evaluate the PEEs as follows: if det$[F_{ij}]>0$, $\Delta h^2=(F^{-1})^{11}$, $\Delta \psi^2=(F^{-1})^{22}$ and $\Delta t_m^2=(F^{-1})^{33}$;
if det$[F_{ij}]=0$, $\Delta h^2=(\mathcal F^{-1})^{11}$, $\Delta \psi^2=(\mathcal F^{-1})^{22}$ and $\Delta t_m^2=+\infty$.

\begin{figure*}
  \centering
  \input{PEEfig.tex}\\
  ~\\
  \caption{PDFs of PEEs for the best-6 strategy in FAST-PTA (including $5$-year IPTA archived data before $t_{I, i}$) with $\tau=1000$ h. 
  The results for Model-1 is shown in the left panel and for Model-2 shown in the right panel. 
  The IPTA archived data is assumed to be observed at a cadence $\Delta t_I=2$ weeks.}\label{Fig:deltah}
\end{figure*}
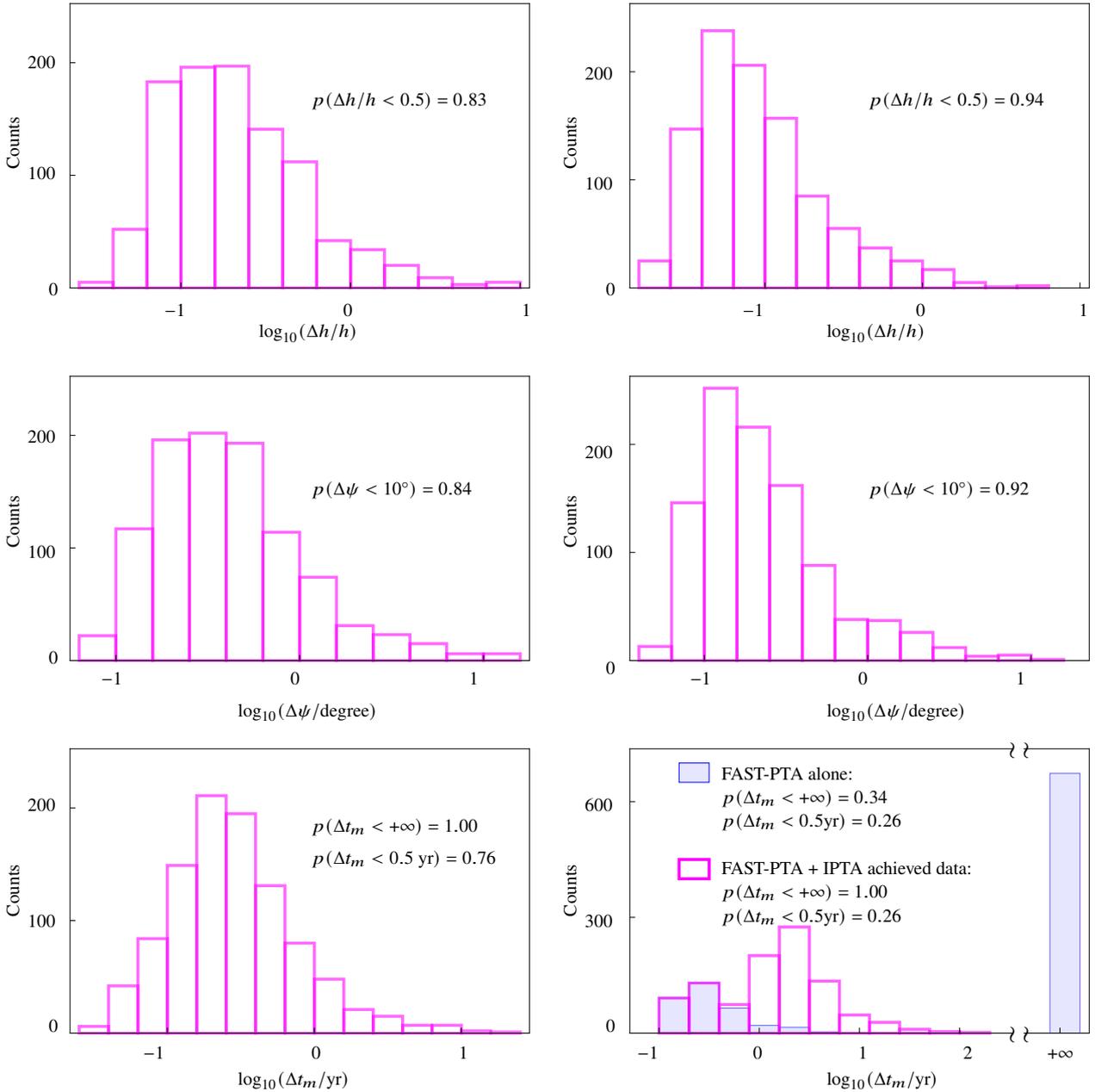

We notice that the probability of $t_m<t_{I, i}$ (i.e. $\Delta t_m=+\infty$) is small for Model-1, but it is not ignorable for Model-2 (as shown in \autoref{Fig:deltah}), if the PTA observation starts at $2022.3$.
To decrease this probability, the the strategy can be  optimized by combining the archived IPTA data before $t_{I, i}$.
Since we have assumed $s_I=0$ at the initial epoch $t_{I, i}$, the timing residual template containing the archived data should be modified to
\begin{align}
\label{eq:waveform-historical}
s_I (t)=
\begin{cases}
 (t-t_{I, 0}) \Theta(t-t_m) F_I   h  & t>t_{I, i}\\
[(t-t_m) \Theta(t-t_m)-(t_{I, 0}-t_m)] F_I   h & t<t_{I, i}
\end{cases}.
\end{align}
We then extend the red noise model in \autoref{Eq:red-noise} to $\hat \sigma_{I, R} \propto |t-t_{I, i}|^{2.3}$, making it apply for the epoch $t<t_{I, i}$.

We illustrate the results of PEEs in \autoref{Fig:deltah}.
Firstly, it is clear that $\Delta h/h$ and $\Delta \psi$
have significant probabilities to achieve the precision levels of $50\%$ and $10^\circ$, respectively,
for both Model-1 and Model-2. 
Secondly, $\Delta t_m$ can achieve the $0.5$-year level with a high confidence for Model-1, but
have a large probability of $\Delta t_m=+\infty$ and low probability of $\Delta t_m<0.5$ yr in Model-2, if the FAST-PTA data are used alone. 
Furthermore, combining the IPTA archived data can
eliminate the probability of $\Delta t_m=+\infty$. 
This indicates that the IPTA archived data is important in measuring $t_m$. 
Additionally, since the IPTA precision is not as good as FAST-PTA, the $5$-year archived data cannot improve the probability of $\Delta t_m<0.5$ yr significantly.
Hence, precisely measuring $\Delta t_m$ is still a challenge for future PTA observations.


\section{Summary}\label{sec:summary}

In this article, the prospects for detecting the GW BWM signal from SDSSJ1430$+$0323 by IPTA and FAST-PTA are investigated.
For this target signal, we present a 
detailed process of designing observation strategies, which includes selecting appropriate MSPs and obtaining the observation epochs for all data points.
Our results show that: (i) FAST-PTA have a considerable probability to detect the BWM signal in $5$ years, with total integration time $\sim 500$ hours;
(ii) precisely measuring the merger time $t_m$ may be a challenge for these PTA observations.

Note that our results are based on some idealized treatments,
e.g. the  over-estimated  telescope parameters $G$ and $T_{\rm sys}$ for FAST.
Furthermore, our analysis of the red noise $\sigma_{I, R}$ is simple.
This may lead to undesired results from the strategies, e.g. some precise MSPs have very short spans ($T \lesssim 1$ year) or their cadences are required to be very high ($\Delta t_I \lesssim 1$ day).
An improvement of the analyses on $\sigma_{I, R}$ may optimize the strategies by breaking the conditions, and worth being studied in the follow-up works.
Moreover, there are MSPs not considered in this work \citep[e.g. see][]{PPTA-DR2, NANOGrav-narrow, NANOGrav-wide}.
It indicates that our results will be updated when including these MSPs in the future.

As to the challenge in measuring $\Delta t_m$,
FAST may have already accumulated precise pulsar timing data before $2022.3$,
for the selected MSPs in \autoref{sec:PEE}.
Combining the FAST archived data is expected to further reduce the measurement error $\Delta t_m$ from our results.
Additionally, besides PTA observations, the GW memory may also trigger signals for the ground-based LIGO and Virgo \citep{Yang-Martynov:2018}.
In particular, if LIGO and Virgo run recently, $t_m$ can be covered in their observation time range, which provides another chance to measure it.


\section*{Acknowledgements}

J. W. C. acknowledges the support from China Postdoctoral Science Foundation under Grant No. 2021M691146.
Y. M. thanks Huan Yang, Yanbei Chen, Ning Jiang, Zhenwei Lv and Qingwen Wu for many helpful discussions on the SDSSJ1430 source and also acknowledges the support from the university start-up fundings of Huazhong University of Science and Technology.
Y. W. gratefully acknowledges support from the National Key Research and Development Program of China (No. 2020YFC2201400), the National Natural Science Foundation of China (NSFC) under Grants No. 11973024, and Guangdong Major Project of Basic and Applied Basic Research (Grant No. 2019B030302001).

\section*{Data Available Statement} 

The data underlying this article will be shared on reasonable request to the corresponding authors.



\bibliographystyle{mnras}
\bibliography{manuscript.bib} 






\bsp	
\label{lastpage}
\end{document}

%% file: msptable.tex
\begin{table*}
\caption{Parameters of 65 IPTA MSPs. Column 2 shows the measured $\sigma_{I, W}$ given by IPTA DR2 \citep{IPTA-DR2}; columns 3-5 give the measured $S_I$ at $1.4$ GHz, $P_I$, and $W_I$ (at $50\%$ intensity), respectively, 
where the data are cited from IPTA DR2 \citep{IPTA-DR2}, EPTA 2016 results \citep{EPTA-2016}, PPTA DR2 \citep{PPTA-DR2}, NANOGrav $12.5$-year results \citep{NANOGrav-narrow, NANOGrav-wide}, the
ATNF Pulsar Catalogue \citep{Manchester_2005} (\href{http://www.atnf.csiro.au/research/pulsar/psrcat/}{http://www.atnf.csiro.au/research/pulsar/psrcat/}) and the references therein;
column 6 illustrates the evaluated $\sigma_{I, W}$ by FAST with $\Delta \tau_I=5$ min, from \autoref{eq:white-noise}; column 7 lists the observed $\dot P_I$ cited from the ATNF Pulsar Catalogue;
columns 8-9 show parameters $\hat \sigma_{I, R, \rm 1 yr}$ and $\delta_I$ in \autoref{Eq:red-noise};
columns 10 shows $T_I^{*}$ given by \autoref{eq:span-limit} for FAST-PTA, and columns 11 shows $T_I^{*}$ for IPTA, assuming that the white noises in column 2 are measured at $\Delta \tau_I=20$ min for all MSPs.
}
\label{Tab:white-noise}
\begin{tabular}{|c|c|c|c|c|c|c|c|c|c|c|}
\hline
~~~~~MSP~~~~~ & $\sigma_{I, W, {\rm IPTA}}$ & $S_{I, 1.4}$ & $P_I$ & $W_{I, 50}$ & $\sigma_{I, W, 5 {\rm min}, {\rm FAST}}$ &   $\dot P_I$ & $\hat \sigma_{I, R, \rm 1 yr}$ & $\delta_I$ & $  T^{*}_{I, {\rm FAST}}$ & $  T^{*}_{I, {\rm IPTA}}$ \\
name & (ns)  & (mJy) & (ms) & (ms) & (ns)  & ($10^{-20}$) & (ns) & ~ & (yr) & (yr)\\
\hline
J0023$+$0923  & $1340$  & $0.32$ & $3.050$ & $-$ & $-$ & $1.14$ & $1.93$ & $0.88$ & $-$ & $26.6$ \\
J0030$+$0451 & $1480$  & $1.12$ & $4.856$ & $0.8$ & $1034$ & $1.02$ & $1.28$ & $0.86$ & $21.1$ & $33.3$  \\
J0034$-$0534 & $4190$ & $0.61$ & $1.877$ & $0.65$ & $2005$ & $0.50$ &  $1.40$ & $0.91$ & $27.0$ & $50.3$ \\
J0218$+$4232  & $7010$ & $0.9$ & $2.323$ & $1.03$ & $2669$ & $7.74$ &  $10.8$ & $0.90$ & $12.6$ & $25.9$ \\
J0340$+$4130 & $5160$ & $0.54$ & $3.299$ & $-$ & $-$ & $0.70$ & $1.23$ & $0.88$ & $-$ & $58.3$ \\
J0613$-$0200 & $1140$ & $2.3$ & $3.062$ & $0.462$ & $366$ & $0.96$ & $1.68$ & $0.88$ & $11.9$ & $26.4$ \\
J0621$+$1002 & $6570$ & $1.9$ & $28.854$ & $0.69$ & $1361$  & $4.73$ &  $1.25$ & $0.79$ & $24.0$ & $64.3$ \\
J0645$+$5158  & $570$  & $0.3$ & $8.853$ & $0.26$ & $477$& $0.49$ & $0.46$ & $0.83$ & $23.5$ & $34.3$ \\
J0751$+$1807 & $3000$ & $3.2$ & $3.479$ & $0.7$ & $553$ & $0.78$ & $1.30$ & $0.88$ & $15.9$ & $44.9$ \\
J1012$+$5307 & $1910$  & $3.2$ & $5.256$ & $0.85$ & $771$ & $1.71$ & $1.82$ & $0.86$ & $15.9$ & $31.9$ \\
J1022$+$1001 & $1970$ & $6.1$ & $16.453$ & $0.972$ & $1443$ & $4.33$ & $1.72$ & $0.81$ & $21.4$ & $33.1$ \\
J1024$-$0719 & $1710$ & $1.5$ & $5.162$ & $0.521$ & $524$ & $1.86$ & $1.97$ & $0.86$ & $13.0$ & $29.4$ \\
J1640$+$2224 & $770$ & $2.0$ & $3.163$ & $0.22$ & $162$  & $0.28$ & $0.61$ & $0.88$ & $13.0$ & $34.6$ \\
J1643$-$1224 & $2550$ & $4.8$ & $4.622$ & $0.314$ & $251$  & $1.85$ & $2.13$ & $0.86$ & $9.1$ & $33.8$ \\
J1713$+$0747 & $210$ & $10.2$ & $4.570$ & $0.11$ & $86$ & $0.85$ & $1.15$ & $0.86$ & $7.5$ & $14.9$ \\
J1738$+$0333 & $1380$ & $0.7$ & $5.850$ & $0.43$ & $582$ & $2.41$ & $2.22$ & $0.85$ & $12.9$ & $25.4$ \\
J1741$+$1351 & $460$  & $0.9$ & $3.747$ & $0.16$ & $148$ & $3.02$ & $3.63$ & $0.87$ & $5.7$ & $12.7$ \\
J1744$-$1134 & $730$ & $3.1$ & $4.075$ & $0.137$ & $103$  & $0.89$ & $1.30$ & $0.87$ & $7.7$ & $24.3$ \\
J1832$-$0836  & $1860$ & $1.18$ & $2.719$ & $0.058$ & $40$ & $0.83$ & $1.62$ & $0.89$ & $4.6$ & $33.2$ \\
J1843$-$1113 & $710$ & $0.1$ & $1.846$ & $0.25$ & $2528$ & $0.96$ & $2.38$ & $0.91$ & $23.7$ & $18.5$ \\
J1853$+$1303 & $1310$ & $0.43$ & $4.092$ & $0.59$ & $1501$  & $0.87$ & $1.26$ & $0.87$ & $24.9$ & $31.8$ \\
J1857$+$0943 & $1160$ & $5.0$ & $5.362$ & $0.518$ & $446$ & $1.78$ & $1.86$ & $0.85$ & $12.4$ & $25.4$  \\
J1903$+$0327 & $2110$  & $1.3$ & $2.150$ & $-$ & $-$ & $1.88$ & $3.68$ & $0.90$ & $-$ & $24.5$ \\
J1910$+$1256 & $1420$ & $0.56$ & $4.984$ & $0.14$ & $157$ & $0.97$ &  $1.20$ & $0.86$ & $9.5$ & $33.6$ \\
J1911$-$1114 & $4300$  & $0.5$ & $3.626$ & $0.18$ & $244$ & $1.40$ & $2.01$ & $0.87$ & $9.2$ & $43.5$ \\
J1911$+$1347 & $1090$ & $0.86$ & $4.626$ & $0.089$ & $79$  & $1.69$ & $1.97$ & $0.86$ & $5.7$ & $24.1$ \\
J1918$-$0642  & $1800$  & $1.36$ & $7.646$ & $0.66$ & $767$ & $2.57$ & $1.94$ & $0.84$ & $15.4$ & $30.2$ \\
J1923$+$2515  & $2250$ & $0.28$ & $3.788$ & $0.38$ & $1187$ & $0.96$ & $1.45$ & $0.87$ & $21.2$ & $37.8$ \\
J1939$+$2134 & $240$ & $13.2$ & $1.558$ & $0.038$ & $17.4$  & $10.5$ & $18.3$ & $0.92$ & $1.1$ & $4.7$ \\
J1944$+$0907 & $2220$ & $2.6$ & $5.185$ & $0.5$ & $446$ & $1.73$ & $1.85$ & $0.86$ & $12.4$ & $33.8$ \\
J1949$+$3106 & $4610$ & $0.2$ & $13.138$ & $-$ & $-$  & $9.39$ & $3.75$ & $0.82$ & $-$ & $34.2$ \\
J1955$+$2908 & $3200$ & $1.1$ & $6.133$ & $0.65$ & $784$  & $2.97$ & $2.54$ & $0.85$ & $13.9$ & $34.5$ \\
J2010$-$1323  & $2530$ & $1.6$ & $5.233$ & $0.28$ & $257$ & $0.48$ & $0.66$ & $0.86$ & $15.3$ & $56.3$ \\
J2017$+$0603 & $720$ & $0.5$ & $2.896$ & $-$ & $-$ & $0.80$ & $1.51$ & $0.88$ & $-$ & $22.6$ \\
J2019$+$2425 & $9640$ & $0.1$ & $3.934$ & $0.3$ & $2210$ & $0.70$  & $1.09$ & $0.87$ & $31.4$ & $80.6$ \\
J2033$+$1734 & $13650$ & $0.3$ & $5.949$ & $0.16$ & $267$ & $1.11$ & $1.18$ & $0.85$ & $12.1$ & $90.5$ \\
J2043$+$1711 & $630$ & $0.21$ & $2.380$ & $-$ & $-$ & $0.52$ & $1.22$ & $0.89$ & $-$ & $23.4$ \\
J2145$-$0750 & $1730$ & $8.9$ & $16.052$ & $0.337$ & $494$ & $2.98$ & $1.30$ & $0.81$ & $15.2$  & $35.4$ \\
J2214$+$3000 & $1670$ & $0.5$ & $3.199$ & $0.23$ & $360$ & $1.47$ & $2.28$ & $0.88$ & $10.4$ & $27.3$ \\
J2229$+$2643 & $4280$ & $0.9$ & $2.978$ & $0.58$ & $887$ & $0.15$ & $0.39$ & $0.88$ & $33.0$ & $88.5$ \\
J2302$+$4442 & $5820$ & $1.2$ & $5.192$ & $0.34$ & $342$ & $1.39$ & $1.56$ & $0.86$ & $11.9$  & $55.4$ \\
J2317$+$1439 & $870$ & $4$ & $3.445$ & $0.4$ & $286$  & $0.24$ & $0.51$ & $0.87$ & $18.0$ & $39.1$ \\
J2322$+$2057 & $6740$ & $0.03$ & $4.808$ & $0.3$ & $6587$ & $0.97$ & $1.23$ & $0.86$ & $47.8$ & $199$ \\
J0437$-$4715  & $110$  & $149$ & $5.757$ & $0.14$ & $-$ & $5.73$ & $4.48$ & $0.85$ & $-$ & $6.2$\\
J0610$-$2100  & $4880$  & $0.4$ & $3.816$ & $0.57$ & $-$ & $1.23$ & $1.76$ & $0.87$ & $-$ & $48.7$\\
J0711$-$6830  & $1440$  & $3.2$ & $5.491$ & $1.09$ & $-$ & $1.49$ & $1.60$ & $0.85$ & $-$ & $29.8$\\
J0900$-$3144  & $3210$  & $3.8$ & $11.11$ & $0.80$ & $-$ & $4.89$ & $2.51$ & $0.82$ & $-$ & $34.8$\\
J0931$-$1902  & $3690$  & $0.84$ & $4.638$ & $0.45$ & $-$ & $0.36$ & $0.57$ & $0.86$ & $-$ & $70.3$\\
J1045$-$4509  & $3190$  & $2.7$ & $7.474$ & $0.84$ & $-$ & $1.77$ & $1.47$ & $0.84$ & $-$ & $43.7$\\
J1455$-$3330  & $4120$  & $1.2$ & $7.987$ & $0.80$ & $-$ & $2.43$ & $1.80$ & $0.84$ & $-$ & $44.8$\\
J1600$-$3053  & $920$  & $2.5$ & $3.598$ & $0.094$ & $-$ & $0.95$ & $1.48$ & $0.87$ & $-$ & $25.4$ \\
J1603$-$7202  & $1580$  & $3.1$ & $14.84$ & $1.21$ & $-$ & $1.56$ & $0.82$ & $0.81$ & $-$ & $41.5$ \\
J1614$-$2230  & $1380$  & $1.11$ & $3.151$ & $0.30$ & $-$ & $0.96$ & $1.65$ & $0.88$ & $-$ & $28.9$ \\
J1721$-$2457  & $12210$  & $0.6$ & $3.497$ & $0.58$ & $-$ & $0.55$ & $0.98$ & $0.87$ & $-$ & $93.5$ \\
J1730$-$2304  & $1570$  & $3.9$ & $8.123$ & $0.965$ & $-$ & $2.02$ & $1.54$ & $0.84$ & $-$ & $31.5$ \\
\hline
\end{tabular}
\end{table*}

\begin{table*}
\leftline{\textbf{Table 1.} --- continued.} 
~\\
\begin{tabular}{|c|c|c|c|c|c|c|c|c|c|c|}
\hline 
~~~~~MSP~~~~~ & $\sigma_{I, W, {\rm IPTA}}$ & $S_{I, 1.4}$ & $P_I$ & $W_{I, 50}$ & $\sigma_{I, W, 5 {\rm min}, {\rm FAST}}$ &   $\dot P_I$ & $\hat \sigma_{I, R, \rm 1 yr}$ & $\delta_I$ & $  T^{*}_{I, {\rm FAST}}$ & $  T^{*}_{I, {\rm IPTA}}$ \\
name & (ns)  & (mJy) & (ms) & (ms) & (ns)  & ($10^{-20}$) & (ns) & ~ & (yr) & (yr)\\
\hline
J1732$-$5049  & $2720$  & $1.3$ & $5.313$ & $0.29$ & $-$ & $1.42$ & $1.56$ & $0.85$ & $-$ & $39.8$ \\
J1747$-$4036  & $4790$  & $1.55$ & $1.646$ & $-$ & $-$ & $1.31$ & $3.31$ & $0.91$ & $-$ & $36.7$ \\
J1751$-$2857  & $2850$  & $0.1$ & $3.915$ & $0.25$ & $-$ & $1.12$ & $1.60$ & $0.87$ & $-$ & $38.4$ \\
J1801$-$1417  & $2760$  & $0.2$ & $3.625$ & $0.60$ & $-$ & $0.53$ & $0.93$ & $0.87$ & $-$ & $50.1$ \\
J1802$-$2124  & $2760$  & $0.8$ & $12.65$ & $0.37$ & $-$ & $7.26$ & $3.13$ & $0.82$ & $-$ & $29.6$ \\
J1804$-$2717  & $3720$  & $0.4$ & $9.343$ & $1.55$ & $-$ & $4.09$ & $2.45$ & $0.83$ & $-$ & $37.4$ \\
J1824$-$2452A  & $570$  & $2.0$ & $3.054$ & $0.97$ & $-$ & $1.62$ & $2.55$ & $0.88$ & $-$ & $16.3$ \\
J1909$-$3744  & $190$  & $2.1$ & $2.947$ & $0.044$ & $-$ & $1.40$ & $2.32$ & $0.88$ & $-$ & $10.5$ \\
J2124$-$3358  & $2890$  & $3.6$ & $4.931$ & $0.52$ & $-$ & $2.06$ & $2.22$ & $0.86$ & $-$ & $35.0$ \\
J2129$-$5721  & $980$  & $1.1$ & $3.726$ & $0.26$ & $-$ & $2.09$ & $2.72$ & $0.87$ & $-$ & $20.0$ \\
\hline
\end{tabular}
\end{table*}

%% file: best_msp.tex
\begin{table*}
\caption{Selected 20 MSPs with largest $p_I$ and their parameters for a $5$-year FAST-PTA (left) and  a $5$-year IPTA (right) observations started at $2022.3$, respectively. The default values of span and initial date are $T_I=5$ yr  and $t_{I, i}=2022.3$. }
\label{tab:strategy}
\begin{tabular}{|c c c c c c|c c c c c c|}
\hline
priority & MSP &  $p_I$ & $T_I$ & $t_{I, i}$   & $\tau_I$  & priority &  MSP & $p_I$ & $T_I$ & $t_{I, i}$   & $\tau_I $ \\
(FAST) & name  & $\overline{{\rm \bf Max}[p_{I}]}$  & (yr) &  &   $\overline{{\rm \bf Min}[\tau_{I}]}$  & (IPTA) & name  & $\overline{{\rm \bf Max}[p_{I}]}$ & (yr) &  &  $\overline{{\rm \bf Min}[\tau_{I}]}$  \\
\hline
1 & ~J1832$-$0836~ & $1$ & $4.6$ & ~$-$~ &  $1$ & 1 & ~J1713$+$0747~ & $1$& $-$ & ~$-$~  & $1$  \\
2 & J1713$+$0747 & $0.57$ & $-$  & $-$  & $1.76$ & 2 & J1909$-$3744 &  $0.40$ & $-$ & $-$  & $2.48$\\
3 & J1911$+$1347 & $0.36$ & $-$  & $-$ & $2.81$  & 3 & J1939$+$2134 & $0.26$ & $4.7$ & $-$ & $3.92$\\
4 & J1744$-$1134 & $0.30$ & $-$  & $-$  & $3.30$ & 4 & J1741$+$1351 & $0.20$ & $-$ & $-$ & $5.11$ \\
5 & J1939$+$2134 & $0.28$ & $1.1$ & $2023.4$  & $3.53$ & 5 & J1640$+$2224 & $0.086$ & $-$ & ~$-$~ &  $11.7$ \\
6 & J1640$+$2224 & $0.20$ & $-$  &  $-$  & $4.96$ & 6 & J1824$-$2452A & $0.072$ & $-$ & ~$-$~  & $13.8$\\
7 & J1741$+$1351 & $0.15$ & $-$  & $-$   & $6.64$ & 7 & J1744$-$1134 & $0.063$ & $-$ & ~$-$~  & $15.9$\\
8 & J1910$+$1256  & $0.12$ & $-$ & $-$ & $8.58$ & 8 & J1843$-$1113 & $0.051$ & $-$ & $-$  & $19.6$\\
9 & J1643$-$1224 & $0.065$ & $-$  & $-$  & $15.5$ & 9 & J0645$+$5158 & $0.050$ & $-$ & $-$  & $19.9$\\
10 & J1911$-$1114 & $0.037$ & $-$ & $-$  & $26.8$ & 10 &J2043$+$1711 & $0.042$ & $-$ & ~$-$~ & $24.0$ \\
11 & J2033$+$1734 & $0.026$  & $-$ & $-$ & $38.5$ & 11 & J1600$-$3053 & $0.040$ & $-$ & ~$-$~  & $24.9$ \\
12 &  J2010$-$1323 & $0.023$ & $-$ & $-$ & $44.4$ & 12 & J2017$+$0603  & $0.034$ & $-$ & ~$-$~ & $29.8$  \\
13 & J1857$+$0943 & $0.015$ & $-$ & $-$  & $64.7$ & 13 & J0437$-$4715 & $0.032$ & $-$ & ~$-$~  & $31.3$ \\
14 & J1738$+$0333 & $0.012$ & $-$ & $-$  & $81.9$ & 14 & J1911$-$1347 & $0.024$ & $-$ & ~$-$~  & $41.9$ \\
15 & J1944$+$0907 & $0.012$ & $-$ & $-$   & $84.9$ & 15 & J1857$+$0943 & $0.022$  & $-$ & ~$-$~ & $45.5$\\
16 & J1024$-$0719 & $0.011$ & $-$ & $-$  & $94.2$ & 16 & J1738$+$0333 & $0.021$ & $-$ & ~$-$~ & $48.0$\\
17 & J2302$+$4442 & $0.0098$ & $-$ & $-$  & $102$ & 17 & J1614$-$2230 & $0.020$ & $-$ & ~$-$~  & $50.2$\\
18 & J2214$+$3000 & $0.0083$ & $-$ & $-$ & $121$ & 18 & J1853$+$1303 & $0.018$ & $-$ & ~$-$~  & $55.5$ \\
19 & J0645$+$5158 & $0.0075$ & $-$ & $-$  & $133$ & 19 & J1910$+$1256 & $0.014$  & $-$ & ~$-$~ & $71.2$\\
20 & J1012$+$5307 & $0.0062$ & $-$ & $-$  & $162$ & 20 & J1730$-$2340 & $0.012$  & $-$ & ~$-$~ & $81.7$ \\
\hline
\end{tabular}
\end{table*}

%% file: flow.tex
\tikzstyle{box}=[rectangle, rounded corners, draw, thick, fill=blue!5 ]
\tikzstyle{box0}=[rectangle, rounded corners, draw, thin, fill=white]
\tikzstyle{box2}=[rectangle, rounded corners, draw, thick, minimum width=300, minimum height=140, fill=blue!5]
\tikzstyle{box3}=[rectangle, rounded corners, draw, dashed, thick, minimum width=220, minimum height=46, fill=green!5]
\tikzstyle{box4}=[rectangle, rounded corners, draw, thin, minimum width=100, minimum height=60, fill=white]
\tikzstyle{dia}=[diamond, draw, thin, aspect=4,  text centered, fill=white]
\tikzstyle{dia2}=[diamond, draw, thin, aspect=2,  text centered, fill=white]
\tikzstyle{arrow} = [thick,->,>=stealth, orange]
\begin{tikzpicture}
\node at (2,-3.5) [box2] {};
\node at (-0.5,-8) [box3] {};
\node (step1) at (-4.5, -2.5) [box] {\textbf{Step 1}};
\node (step2-5) at (-4.5,-4.5) [box] {\textbf{Steps 2-5}};
\node  at (-5.6,-3.5) [] {$\Delta \tau_I=5$ min};
\node  at (-5.5,-5.4) [] {input};
\node  at (-5.5,-5.7) [] {$N_P$ and $\tau$};
\node  at (-0.5,-1.5) [] {\textbf{Step 6}:};
\node (step6-1) at (4.5,-2.5) [box0] { $\Delta \tau_I={\rm \bf Min} \left[5~{\rm min}\times \left(\frac{T_I^*}{T_I}\right)^{4.6}, 1~{\rm hour} \right]$};
\node (step6-2)  at (4,-3.8) [box0] { $\Delta t_I=  {\rm \bf Max} \left[ 1~{\rm day} \, , \, {\rm \bf Min} \left(\frac{T_I}{\tau_I/\Delta \tau_I-1}, 4~{\rm weeks} \right) \right]   $}; 
\node (step6-3) at (4,-5) [box0] { $\Delta \tau_I = \frac{\tau_I}{T_I/\Delta t_I+1}$ };
\node (result-1) at (-2.5,-8) [box0] {$p_I$, $\tau_I$, $t_{I, i}$ };
\node (result-2) at (1.5,-8) [box0] {$T_I$, $\Delta t_I$, $\Delta \tau_I$ };
\node at (-0.5, -7.7) [] {\textbf{Results}:};
\node (step6-0) at (-1,-2.5) [dia] {~~~~~~~~~~~~~~~~~~~~~~~~~~~~~~~~~~~~~~~~~~~~};
\node  at (1.5,-2.3) [] { Yes}; 
\node at (-1,-2.5) [] {$\tau_I \geq \left( \frac{T_I}{4~{\rm weeks}}+1 \right)\times 5~{\rm min}$};
\node (step6-0-2) at (-1,-4.5) [box4] {};
\node  at (-1,-4) [] {$\Delta \tau_I=5$ min};
\node  at (-1,-4.5) [] {$\Delta t_I=4$ weeks};
\node  at (-1,-5) [] {$T_I= \left(\frac{\tau_I}{5~{\rm min}}-1 \right) \times 4~{\rm weeks}$};
\draw[arrow](step1) -- (step2-5);
\draw[arrow](4, -3) -- (step6-2);
\draw[arrow](step6-2) -- (step6-3);
\draw[arrow](step2-5)--(-4.5,-6.25)--(-3.6,-6.25)--(-3.6,-2.5)-- (step6-0);
\draw[arrow](-4.1,-6.25)--(-4.1,-8)--(result-1);
\draw[arrow](-4.5,-3.5)--(-4.8,-3.5);
\draw[arrow](-4.8,-5.55)--(-4.5,-5.55);
\draw[arrow](step6-0)--(step6-1);
\draw[arrow](step6-0)--(step6-0-2);
\draw[arrow](1.5,-6.25)--(result-2);
\draw[-,thick,orange](step6-0-2)--(-1,-6.25)--(4,-6.25)--(step6-3);
\end{tikzpicture}

%% file: duty-cycle.tex
  \begin{overpic}[width=0.9\textwidth]{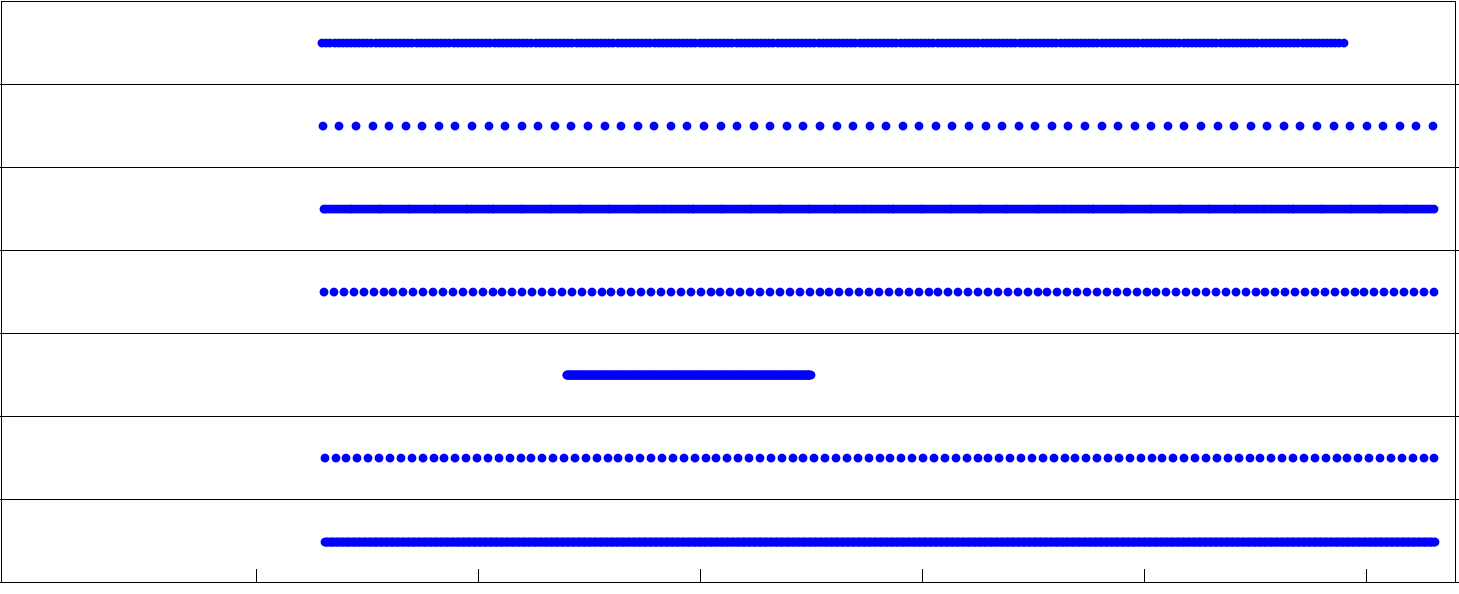}
  \put(45,-5){Date (year)}
  \put(16,-2){$2022$} \put(31,-2){$2023$} \put(46,-2){$2024$} \put(61,-2){$2025$} \put(76,-2){$2026$} \put(91,-2){$2027$} \put(1,37){J1832$-$0836} 
  \put(10.5,39){\footnotesize $\tau_I=20.7$ hours} 
  \put(10.5,37.25){\footnotesize $\Delta t_I=7$ days}
  \put(10.5,35.5){\footnotesize $\Delta \tau_I=5.2$ min}
  \put(1,31.5){J1713$+$0747} 
  \put(10.5,33.5){\footnotesize $\tau_I=36.1$ hours} 
  \put(10.5,31.75){\footnotesize $\Delta t_I=27$ days}   
  \put(10.5,30){\footnotesize $\Delta \tau_I=31.4$ min}
  \put(1,26){J1911$+$1347} 
  \put(10.5,27.75){\footnotesize $\tau_I=58.4$ hours}
  \put(10.5,25.75){\footnotesize $\Delta t_I=5$ days}
  \put(10.5,24){\footnotesize $\Delta \tau_I=9.6$ min}
 \put(1,20){J1744$-$1134}
 \put(10.5,21.75){\footnotesize $\tau_I=68.2$ hours}
 \put(10.5,20){\footnotesize $\Delta t_I=16$ days}
 \put(10.5,18.25){\footnotesize $\Delta \tau_I=35.6$ min}
 \put(1,14){J1939$+$2134}
 \put(10.5,16){\footnotesize $\tau_I=73.4$ hours}
 \put(10.5,14.25){\footnotesize $\Delta t_I=1$ day}
 \put(10.5,12.5){\footnotesize $\Delta \tau_I=10.9$ min}
  \put(1,9){J1640$+$2224}
 \put(10.5,10.75){\footnotesize $\tau_I=101.4$ hours}
 \put(10.5,9.0){\footnotesize $\Delta t_I=18$ days}
 \put(10.5,7.25){\footnotesize $\Delta \tau_I=60.1$ min}
 \put(1,3){J1640$+$2224}
 \put(10.5,5){\footnotesize $\tau_I=138.2$ hours}
 \put(10.5,3.25){\footnotesize $\Delta t_I=2$ days}
 \put(10.5,1.5){\footnotesize $\Delta \tau_I=9.1$ min}
  \end{overpic}

%% file: statistical_signature.tex
\begin{table*}
    \caption{Statistical signatures for a $5$-year FAST-PTA observation (upper) and  a $5$-year IPTA observation (lower). 
    The results in bold face denote the largest $p(\rho>8)$ among different strategies at a fixed $\tau$.   }
    \label{tab:PDF}
    FAST-PTA results for Model-1 (left) and Model-2 (right)\\
    \begin{tabular}{|c||c|c|c||c|c|c|}
    \hline
     strategy   &  $\sqrt{\langle\rho^2\rangle}$ & $p(\rho>8)$ & $p(\lambda<0.5)$  &  $\sqrt{\langle\rho^2\rangle}$ & $p(\rho>8)$ & $p(\lambda<0.5)$ \\
    \hline
    $\tau$ / hours  &  $~\,500~~\big|~~1000~~\big|~~2000$ &  $~\,500~~\big|~~1000~~\big|~~2000$ &  $~\,500~~\big|~~1000~~\big|~~2000$ &   $~\,500~~\big|~~1000~~\big|~~2000$ &  $~\,500~~\big|~~1000~~\big|~~2000$ &  $~\,500~~\big|~~1000~~\big|~~2000$ \\
    \hline 
   best-$1$   & $15.4~~\big|~~ 21.6 ~~\big|~~ 28.0 $ & $0.40 ~~\big|~~ 0.43 ~~\big|~~ 0.48 $ & $ 0.00 ~~\big|~~0.00~~\big|~~ 0.00 $  & $ 26.2 ~~\big|~~  34.3 ~~\big|~~ 48.3 $ & $ 0.66 ~~\big|~~ 0.44 ~~\big|~~ 0.68 $ & $ 0.00 ~~\big|~~ 0.00 ~~\big|~~ 0.00$ \\
     best-$2$   & $13.5 ~~\big|~~ 19.7 ~~\big|~~ 25.6 $ & $0.44 ~~\big|~~ 0.56 ~~\big|~~ 0.64 $ & $ 0.00 ~~\big|~~0.00~~\big|~~ 0.00 $  & $ 23.9 ~~\big|~~  33.0 ~~\big|~~ 43.2 $ & $ 0.70 ~~\big|~~ 0.74 ~~\big|~~ 0.82 $ & $ 0.00 ~~\big|~~ 0.00 ~~\big|~~ 0.00$ \\
     best-$3$   & $12.6 ~~\big|~~ 17.3 ~~\big|~~ 22.7 $ & ${\bf 0.48} ~~\big|~~ 0.59 ~~\big|~~ 0.70 $ & $ 0.12 ~~\big|~~ 0.14 ~~\big|~~ 0.10 $  & $ 20.4 ~~\big|~~  28.9 ~~\big|~~ 38.2 $ & $ {\bf 0.71} ~~\big|~~ 0.80 ~~\big|~~ 0.85 $ & $ 0.14 ~~\big|~~ 0.15 ~~\big|~~ 0.14$ \\
     best-$4$   & $11.4 ~~\big|~~ 16.6 ~~\big|~~ 22.3 $ & $0.46 ~~\big|~~ {\bf 0.61} ~~\big|~~ {\bf 0.73} $ & $ 0.31 ~~\big|~~0.30~~\big|~~ 0.26 $  & $ 17.9 ~~\big|~~  25.9 ~~\big|~~ 36.0 $ & $ {\bf 0.71} ~~\big|~~ {\bf 0.82} ~~\big|~~ {\bf 0.88} $ & $ 0.34 ~~\big|~~ 0.35 ~~\big|~~ 0.35$ \\
     best-$5$   & $11.3 ~~\big|~~ 15.6~~\big|~~ 21.6 $ & $0.46  ~~\big|~~ 0.60 ~~\big|~~ 0.71 $ & $ 0.38 ~~\big|~~0.38 ~~\big|~~ 0.34 $  & $ 16.5 ~~\big|~~  23.8 ~~\big|~~ 33.1 $ & $ {\bf 0.71} ~~\big|~~ {\bf 0.82} ~~\big|~~ 0.87 $ & $ 0.48 ~~\big|~~ 0.47 ~~\big|~~ 0.47$ \\ 
     best-$6$   & $10.2 ~~\big|~~ 14.4 ~~\big|~~ 19.0 $ & $0.42 ~~\big|~~ 0.57 ~~\big|~~ 0.70 $ & $ 0.52 ~~\big|~~0.53~~\big|~~ 0.54 $  & $ 15.7 ~~\big|~~  22.4 ~~\big|~~ 31.0 $ & $ 0.69 ~~\big|~~ 0.80 ~~\big|~~ {\bf 0.88} $ & $ 0.66 ~~\big|~~ 0.63 ~~\big|~~ 0.66$ \\
     best-$7$   & $9.40 ~~\big|~~ 12.7 ~~\big|~~ 18.2 $ & $0.39 ~~\big|~~ 0.53 ~~\big|~~ 0.67 $ & $ 0.68 ~~\big|~~0.66 ~~\big|~~ 0.65 $  & $ 14.6 ~~\big|~~  20.6 ~~\big|~~ 28.8 $ & $ 0.65 ~~\big|~~ 0.76 ~~\big|~~ 0.85 $ & $ 0.78 ~~\big|~~ 0.79 ~~\big|~~ 0.77$ \\
     best-$8$   & $8.61 ~~\big|~~ 12.0 ~~\big|~~ 17.9 $ & $0.33 ~~\big|~~ 0.51 ~~\big|~~ 0.66 $ & $ 0.76 ~~\big|~~ 0.77 ~~\big|~~ 0.77 $  & $ 13.3 ~~\big|~~  19.0 ~~\big|~~ 26.8 $ & $ 0.60 ~~\big|~~ 0.76 ~~\big|~~ 0.84 $ & $ 0.84 ~~\big|~~ 0.85 ~~\big|~~ 0.84$ \\
     best-$9$   & $7.40 ~~\big|~~ 10.4 ~~\big|~~ 14.7 $ & $0.28 ~~\big|~~ 0.45 ~~\big|~~ 0.62 $ & $ 0.84 ~~\big|~~0.86~~\big|~~ 0.84 $  & $ 11.9 ~~\big|~~  16.0 ~~\big|~~ 22.4 $ & $ 0.59 ~~\big|~~ 0.72 ~~\big|~~ 0.81 $ & $ 0.94 ~~\big|~~ 0.93 ~~\big|~~ 0.92$ \\
     best-$10$   & $6.38 ~~\big|~~ 8.96 ~~\big|~~ 12.5 $ & $0.21 ~~\big|~~ 0.38 ~~\big|~~ 0.54 $ & $ 0.88 ~~\big|~~0.90~~\big|~~ 0.91 $  & $ 9.80 ~~\big|~~  13.3 ~~\big|~~ 19.6 $ & $ 0.49 ~~\big|~~ 0.65 ~~\big|~~ 0.77 $ & $ 0.97 ~~\big|~~ 0.97 ~~\big|~~ 0.97$ \\     
     best-$15$   & $3.18 ~~\big|~~ 4.66 ~~\big|~~ 6.53 $ & $0.01 ~~\big|~~ 0.08 ~~\big|~~ 0.22 $ & $ 0.98 ~~\big|~~0.99~~\big|~~ 0.99 $  & $ 4.76 ~~\big|~~  7.49 ~~\big|~~ 10.2 $ & $ 0.08 ~~\big|~~ 0.34 ~~\big|~~ 0.50 $ & $ 1.00 ~~\big|~~ 1.00 ~~\big|~~ 1.00$ \\
    \hline 
    \end{tabular}
    ~\\
    ~\\
    IPTA results for Model-1 (left) and Model-2 (right).\\
    \begin{tabular}{|c||c|c|c||c|c|c|}
    \hline
     strategy   & $\sqrt{\langle\rho^2\rangle}$ & $p(\rho>8)$ & $p(\lambda<0.5)$  &  $\sqrt{\langle\rho^2\rangle}$ & $p(\rho>8)$ & $p(\lambda<0.5)$ \\
    \hline
    $\tau$ / hours  &  $~\,500~~\big|~~1000~~\big|~~2000$ &  $~\,500~~\big|~~1000~~\big|~~2000$ &  $~\,500~~\big|~~1000~~\big|~~2000$ &   $~\,500~~\big|~~1000~~\big|~~2000$ &  $~\,500~~\big|~~1000~~\big|~~2000$ &  $~\,500~~\big|~~1000~~\big|~~2000$ \\
    \hline     
    best-1   & $ 2.60 ~~\big|~~ 3.52 ~~\big|~~ 4.39  $ & $  {\bf 0.01 ~~\big|~~0.05~~\big|~~ 0.09}  $ & $ 0.00 ~~\big|~~0.00~~\big|~~ 0.00 $ &  $ 4.07 ~~\big|~~5.34~~\big|~~ 7.16  $ & $  \bf{0.07 ~~\big|~~0.14~~\big|~~ 0.22}  $ & $ 0.00 ~~\big|~~0.00~~\big|~~ 0.00 $ \\
    best-2   & $ 1.91 ~~\big|~~ 2.66 ~~\big|~~ 3.53 $ & $ 0.00 ~~\big|~~ 0.01 ~~\big|~~ 0.04 $ & $ 0.00 ~~\big|~~ 0.00 ~~\big|~~ 0.00 $ &  $ 3.05 ~~\big|~~ 4.04 ~~\big|~~ 5.50 $ & $ 0.01 ~~\big|~~ 0.05 ~~\big|~~ 0.15 $ & $ 0.00 ~~\big|~~ 0.00 ~~\big|~~ 0.00 $ \\
    best-3   & $ 1.52 ~~\big|~~ 2.04 ~~\big|~~ 2.76 $ & $ 0.00 ~~\big|~~ 0.00 ~~\big|~~ 0.01 $ & $ 0.04 ~~\big|~~ 0.04 ~~\big|~~ 0.04 $ &  $ 2.51 ~~\big|~~ 3.56 ~~\big|~~ 4.43 $ & $ 0.00 ~~\big|~~ 0.01 ~~\big|~~ 0.07 $ & $ 0.05 ~~\big|~~ 0.05 ~~\big|~~ 0.05 $ \\
    \hline
\end{tabular}
\end{table*}

%% file: PEEfig.tex
  \begin{overpic}[width=0.9\textwidth]{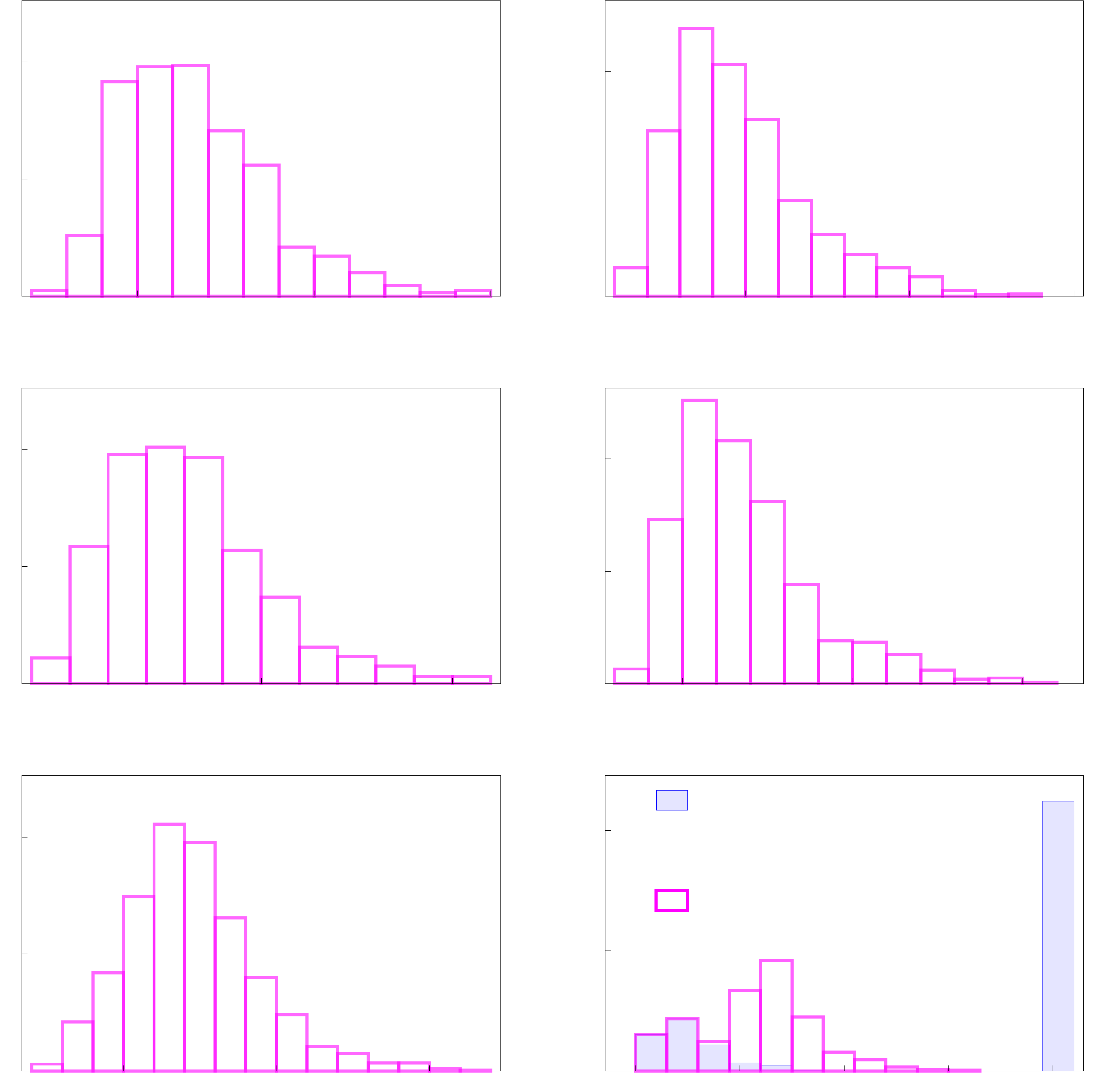}
  \put(11,70){$-1$} \put(28,70){$0$} \put(45,70){$1$} 
  \put(20,68){$\log_{10}(\Delta h/h)$} 
 \put(0,72){$0$}  \put(-2,83){$100$} \put(-2,93.5){$200$}
 \put(-4,84){\begin{sideways} Counts \end{sideways}}
 \put(25,90){$p(\Delta h/h<0.5)=0.83$}
 
 \put(18,32){$\log_{10}(\Delta \psi/ {\rm degree})$}
 \put(5,35){$-1$} \put(23,35){$0$} \put(40,35){$1$}
 \put(0,37){$0$}  \put(-2,47){$100$} \put(-2,58){$200$} 
 \put(-4,48){\begin{sideways} Counts \end{sideways}}
 \put(25,53){$p(\Delta \psi<10^\circ)=0.84$}
 
 \put(18,-3){$\log_{10}(\Delta t_m/ {\rm yr})$}
 \put(9,-0.5){$-1$} \put(25,-0.5){$0$} \put(39,-0.5){$1$}
 \put(0,2){$0$}  \put(-2,12){$100$} \put(-2,23){$200$} 
 \put(-4,13){\begin{sideways} Counts \end{sideways}}
 \put(25,21){$p(\Delta t_m<+\infty)=1.00$}
 \put(25,18){$p(\Delta t_m<0.5~{\rm yr})=0.76$}
 
 \put(66,70){$-1$} \put(82.5,70){$0$} \put(98,70){$1$} 
 \put(74,68){$\log_{10}(\Delta h/h)$} 
 \put(53,72){$0$}  \put(51,82){$100$} \put(51,92.5){$200$}
 \put(49,84){\begin{sideways} Counts \end{sideways}} 
 \put(78,90){$p(\Delta h/h<0.5)=0.94$}
 
 \put(61,35){$-1$} \put(77,35){$0$} \put(92,35){$1$} 
 \put(74,32){$\log_{10}(\Delta \psi/ {\rm degree})$} 
 \put(53,36){$0$}  \put(51,47){$100$} \put(51,56.5){$200$}
 \put(49,48){\begin{sideways} Counts \end{sideways}}
 \put(78,53){$p(\Delta \psi<10^\circ)=0.92$}
 
 \put(74,-3){$\log_{10}(\Delta t_m/ {\rm yr})$}
 \put(56,-0.5){$-1$}  \put(67,-0.5){$0$} \put(77,-0.5){$1$} \put(87,-0.5){$2$} \put(95,-0.5){$+\infty$}
 \put(53,2){$0$}  \put(51,12){$300$} \put(51,23){$600$} 
 \put(49,13){\begin{sideways} Counts \end{sideways}}
 \put(64,26){FAST-PTA alone:}
 \put(64,23.75){$p(\Delta t_m<+\infty)=0.34$}
 \put(64,21.5){$p(\Delta t_m<0.5{\rm yr})=0.26$}
 \put(64,17){FAST-PTA + IPTA achieved data:} 
 \put(64,14.75){$p(\Delta t_m<+\infty)=1.00$}
 \put(64,12.5){$p(\Delta t_m<0.5{\rm yr})=0.26$} 
 
 \put(91.6,1.2){\normalsize \textcolor{white}{$\blacksquare$} }
 \put(91.3, 1){\Large $\wr$}  \put(92.6, 1){\Large $\wr$}
 \put(91.6,28){\normalsize \textcolor{white}{$\blacksquare$} }
 \put(91.3, 28){\Large $\wr$}  \put(92.6, 28){\Large $\wr$} 
 \end{overpic}